\documentclass[journal,10pt]{IEEEtran}

\normalsize
\usepackage{cite}
\usepackage{flushend}
\ifCLASSINFOpdf
  \usepackage[pdftex]{graphicx}

\else

\fi

\usepackage{color}

\begin{document}

\title{A Novel Relay Selection Strategy of Cooperative Network Impaired by Bursty Impulsive Noise}

\author{Md Sahabul Alam,~\IEEEmembership{Student Member,~IEEE,}
        ~Georges Kaddoum,~\IEEEmembership{Member,~IEEE},
        and~Basile L. Agba,~\IEEEmembership{Senior Member,~IEEE}
\thanks{M. S. Alam and G. Kaddoum are with the Electrical Engineering Department,
ETS, University of Quebec, Montreal, QC H3C 1K3, Canada (e-mail:md-sahabul.alam.1@ens.etsmtl.ca).}
\thanks{B. L. Agba is with the Hydro-Quebec Research Institute, Varennes, QC J3X
1S1, Canada.}}

\markboth{This is the author version of the paper that has been accepted for publication in IEEE Transactions on Vehicular Technology}%
{Submitted paper}

\maketitle

\begin{abstract}
Best relay selection (BRS) is crucial in enhancing the performance of cooperative networks. In contrast to most previous works, where the guidelines for BRS are limited to Gaussian noise, in this article, we propose a novel relay selection protocol for a decode-and-forward cooperative network taking into account the bursty impulsive noise (IN). The proposed protocol chooses the $N$'th best relay considering both the channel gains and the states of the IN of the source-relay and relay-destination links. For this scheme, to obtain the state of IN, we propose a state detection algorithm using maximum a posteriori (MAP) detection. To analyze the performance of the proposed protocol, we first derive closed-form expressions for the probability density function (PDF) of the received signal-to-noise ratio assuming all the relays know the state of IN perfectly (genie-condition). Then, these PDFs are used to derive closed-form expressions for the bit error rate (BER) and the outage probability. Finally, we also derive the asymptotic BER and outage expressions to quantify the diversity benefits. We show that the proposed MAP-based $N$'th BRS protocol attains the derived genie-aided analytical results and outperforms the conventional relay selection protocol, optimized for the Gaussian case, and which does not take into account the IN memory.
\end{abstract}


\IEEEpeerreviewmaketitle
\section{Introduction}
For over a decade, cooperative relaying (CR) has been deemed efficient for reliable transmission over fading and interference channels \cite{nosratinia2004cooperative,laneman2004cooperative,laneman2003distributed}. In addition to many other wireless applications, it is specially attractive for wireless sensor network applications, where the sensor nodes may not be able to afford multiple antennas, because of many constraints including their size, cost, power, etc. In particular, opportunistic relaying, where the BRS is performed between the available relays, is an efficient approach to improve the performance of CR as it makes efficient use of the system resources \cite{bletsas2006simple,ibrahim2008cooperative}. Also, the system complexity and the  synchronization requirements are relaxed through opportunistic relaying, compared to other CR schemes where all relays transmit simultaneously or sequentially over orthogonal channels \cite{bletsas2006simple,ibrahim2008cooperative,fareed2009relay,tourki2013new}. Therefore, the techniques and analysis of BRS have received considerable attention in the literature.

In this regard, the authors in \cite{bletsas2006simple} have proposed a BRS technique, where out of all the available relays, a subset of $M$ relays, possessing error-free detection of the source transmission, are first selected. The best relay is then picked from the subset based on the minimum or the harmonic mean of the source-relay ($SR$) and relay-destination ($RD$) channel gains. It is shown that the proposed scheme exhibits the same performance as obtained in the case where all the relays transmit simultaneously through space-time coding \cite{laneman2003distributed}. Ibrahim \textit{et al.} \cite{ibrahim2008cooperative} have introduced another BRS criterion where the best relay is the one that has the maximum value of the instantaneous scaled harmonic mean function of its $SR$ and $RD$ channel gains. The novelty of this protocol relies on the fact that the relay is not required to forward the source information if the direct link from the source to the destination is of high quality. Since a cooperation is not always taking place, this new scheme achieves higher bandwidth efficiency while the full diversity is guaranteed. Fareed \textit{et al.} \cite{fareed2009relay} have presented another BRS method, with a low implementation complexity, requiring neither error detection methods at the relay nodes \cite{bletsas2006simple} nor feedback information at the source \cite{ibrahim2008cooperative}. For this scheme, based on the minimum of the $SR$ and $RD$ links' signal-to-noise ratios (SNRs), the best relay is chosen at the destination node and it is permitted to transmit only if the minimum of its $SR$ and $RD$ links' SNRs is higher than the direct link SNR. Their obtained results demonstrate that the proposed error-prone BRS method is able to extract the full diversity. The authors in \cite{tourki2013new} have investigated an opportunistic regenerative relaying scheme, where similar to \cite{fareed2009relay}, it is assumed that there might be a possible error propagation. To determine the effect of erroneously detected data at the best relay, in their work, they have derived the exact statistics of each hop. Finally, their analyses have been validated through simulations. The authors in \cite{ikki2010performance} have considered the performance analysis of the $N$'th BRS scheme for both decode-and-forward (DF) and amplify-and-forward (AF) CR systems. Their obtained results show that for the special case where $N=1$, the performance of this scheme coincides with the results available in the literature for the BRS under similar circumstances. The authors in \cite{al2018asymptotic} have generalized the asymptotic analysis of an $N$'th BRS problem using extreme value theory for various fading models commonly used to characterize wireless channels. Also, the selection of $N$'th best relay for cognitive DF relay networks and cooperative energy harvesting DF relay networks have been considered in \cite{zhang2015performance,al2019asymptotic} and \cite{zhang2018secrecy}, respectively. The theory of order statistics \cite{davidorder} has been considered as a powerful tool to analyze these performances.

Although instructive, all of the above performance analyses for BRS protocols have been carried out under the assumption of additive white Gaussian noise (AWGN) only. In practice, the noise characteristics usually observed in many environments are inherently impulsive \cite{middleton1977statistical,sacuto2014wide,zimmermann2002analysis,ndomarkov,shongwe2015study,agba2019impulsive,blackard1993measurements,cheffena2012industrial,asiyo2017analysis,bai2017discrete}. For instance, in power substations, due to partial discharge and switching effects, IN with a bursty behavior is generated from the substation equipment \cite{sacuto2014wide,ndomarkov,shongwe2015study,agba2019impulsive}. In addition to substation environments, bursty impulsive noise is also observed in indoor wireless networks \cite{blackard1993measurements}, industrial wireless sensor networks \cite{cheffena2012industrial}, power line communication (PLC) networks \cite{zimmermann2002analysis,asiyo2017analysis}, and digital subscriber loop (DSL) networks \cite{bai2017discrete}. This article is mainly motivated by this kind of situation where the noise exhibits significant bursty impulsive behavior. The performance of BRS protocols in IN and interference limited environments has barely been considered in the literature. The authors in \cite{qian2018performance} have considered the performance analysis of BRS for DF relay-based PLC systems. Although, Bernoulli-Gaussian model is considered to take into account the combined effects of background Gaussian noise and impulsive noise for deriving the cumulative distribution function (CDF) of the received SNR, the BRS is performed based on the standard \textit{max-min} criterion optimized for AWGN channel and the effect of impulsive noise is not considered in the relay selection process. The extension of conventional optimal \textit{max-min} BRS criterion for interference limited environments, in case of AF relaying strategy, has been investigated in \cite{krikidis2009max}. It is shown that the conventional BRS criterion becomes inefficient under this scenario since the presence of interference modifies the \textit{max-min} BRS statistics. While \cite{krikidis2009max} have considered various BRS protocols for CR in the presence of Gaussian interference, the authors in \cite{ahmed2012relay} have investigated the performance of the BRS and partial BRS protocols impaired by generic noise and interference. Through the derived asymptotic error rate expressions, it is apparent that in contrast to the Gaussian case, the performance of BRS in generic noise depends on the noise moments.

However, the analysis of \cite{krikidis2009max} and \cite{ahmed2012relay} assume that the interfering signals are manifested throughout the transmission and lack the flexibility to deal with the presence or absence of IN and its bursty behavior. In this vein, the authors in \cite{alam2016performance} have considered the performance analysis of a single-relay DF CR scheme over Rayleigh faded bursty IN channels and have proposed an optimal receiver structure that utilizes the MAP detection criterion. It is shown that the performance of such channels improve with the utilization of noise memory at the receiver side through MAP detection, and converges to the derived lower bound: the ultimate performance limit of the same channel obtained under the assumption that perfect noise state information is available at the receiver. In \cite{alam2016effect}, the performance of the single-relay scheme is extended to the multi-relay scenario where all the relays transmit sequentially over orthogonal channels \cite{laneman2004cooperative}. It is shown that as in \cite{alam2016performance}, the MAP receiver also achieves the lower bound drawn for the multi-relay DF CR scheme, and performs significantly better than the conventional schemes. The performance of BRS protocols in bursty IN environments have been investigated in \cite{alam2016relay}. It is assumed that out of all the available $M$ relays, a subset of $N$ relays, not affected by IN are selected first and the best relay is chosen among them based on the optimal \textit{max-min} criterion. Although, the scheme has shown to offer considerable performance improvement in comparison to the BRS strategy optimized for AWGN channels, we note that the achievable potential gain of that scheme is rather limited since the best relay is selected among a subset. In addition to that the analysis in \cite{alam2016relay} is limited to the BER performance only for the finite SNR and, since it is assumed that the selected relay is never affected by the IN, the paper used the available SNR PDFs for AWGN to derive the BER.

In this article, we investigate the performance of BRS protocols for a DF CR scheme over Rayleigh fading channels subject to bursty IN where BRS is performed among all the available relays. This work is an extension of \cite{alam2016relay}. Here, in addition to BER, the analysis also includes outage probability and derive closed-form and asymptotic performances for the proposed scenario. To address the bursty behavior of IN samples, we consider a two-state Markov-Gaussian (TSMG) process \cite{fertonani2009reliable}. A TSMG process is a simple and effective way to model the time-correlation among the noise samples \cite{fertonani2009reliable,mitra2010convolutionally}. Also, we consider the realistic scenario of a \textit{fixed DF CR} \cite{fareed2009relay,tourki2013new}, which does not require any error detection and correction at the relay nodes and hence decoding errors might be propagated from the selected relay.

The contributions of this work are summarized as follows.
\begin{itemize}
\item We propose a novel relay selection protocol called $N$'th BRS, based on both the channel gains of the $SR$ and $RD$ links, and the states of IN affecting these links. To obtain the IN state, we propose a MAP-based state detection algorithm \cite{bahl1974optimal}. The objective of considering MAP is to exploit the noise memory in the state detection process.
\item To validate the performance of the proposed protocol, we derive novel closed-form expressions for the PDF of the received SNR at the selected relay and at the destination assuming all the relays know the state of IN perfectly (genie-condition). These PDFs are used to derive closed-form expressions for the BER using BPSK modulation and the outage probability.
\item We further derive the asymptotic BER and outage expressions as these are useful for quick evaluation of the performance and quantify the achievable diversity order.
\end{itemize}
We show that the proposed MAP-based $N$'th BRS attains the derived analytical results for genie-condition and significantly outperforms the conventional relay selection protocol, optimized for AWGN environments, and which does not take into account the noise memory. In addition, it is revealed that, in the different SNR regions, the different relay selection protocols present different diversity orders under similar circumstances and the proposed MAP-based $N$'th BRS protocol achieves the full diversity order in high SNR regions.

The rest of the paper is organized as follows: Section \ref{system_model} introduces the system model. In Section \ref{relay_techniques}, we provide an overview of the relay selection protocols. In Section \ref{ber_analysis} and \ref{outage_analysis}, we provide the performance analysis of the proposed relay selection protocol in terms of BER and outage probability, respectively, and Section \ref{asymptotic_analysis} derives the same performances for high SNR scenarios. Section \ref{N_results} shows the numerical results and finally, Section \ref{conclusion} concludes this work.
\section{System Model}\label{system_model}
\begin{figure}[!t]
  \centering
  \includegraphics[scale=0.55]{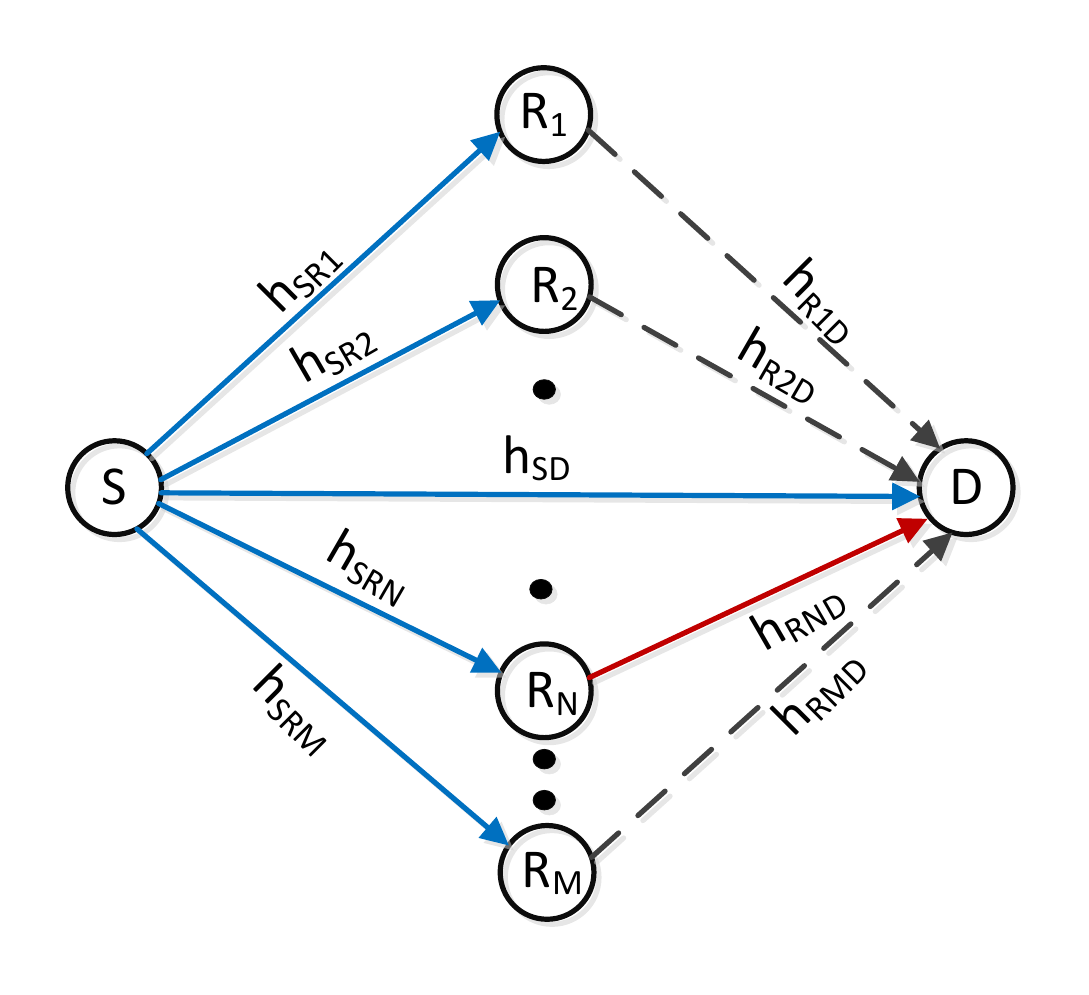}\\
  \caption{Illustration of the considered DF CR with the $N$'th best relay selection.}\label{coop_Mrelays}
\end{figure}
We consider a DF cooperative network where $M$ relays assist the data transmission between the source-destination ($SD$) pair, as shown in Fig~\ref{coop_Mrelays}. We assume that all node terminals have single transmit/receive antennas and share a single communication channel. Also, all nodes are assumed to operate in half-duplex mode. For CR, the transmission is organized in two-time slots. In the first-time slot, the source transmits the data to the destination and the relays. In the second-time slot, the relays form a competition (detailed in section \ref{relay_techniques}) and only the selected relay decodes the message received from the source and forwards it to the destination including possible errors. In our study, during this time, the source remains silent. The destination then combines the noisy sequences received from the source and the selected relay to recover the source information. Although the error propagation problem in this protocol could be resolved by incorporating cyclic redundancy check (CRC) at the relays, we note that this is bandwidth-consuming \cite{wang2007high} and, since CRC checking is usually performed at the MAC layer, it induces excessive signaling overhead. To avoid this, we consider a more general case where there might be a decoding error propagation from the selected relay.
\subsection{Signal Model}
In the first-time slot of the considered CR system, the source $S$ generates a binary information frame of size $K$ $(b_0,b_1,\ldots,b_{K-1})$, mapped into a BPSK modulated sequence $(x_{S,0},x_{S,1},\ldots,x_{S,K-1})$, and broadcasted to the destination and $M$ relay nodes. The signals received at relay $R_m$, $R_m\in\{R_1,R_2,\ldots,R_M\}$ and $D$ at each time epoch $k$, $k=0,1,\ldots,K-1$ can be written, respectively, as
\begin{eqnarray}
y_{{SR_m,}k}&=&\sqrt{P_S}h_{SR_m,k} x_{S,k}+n_{{SR_m,}k},\\
y_{{SD,}k}&=&\sqrt{P_S}h_{SD,k} x_{S,k}+n_{{SD,}k},
\end{eqnarray}
where $P_S$ is the average source transmission power per symbol, $x_{S,k}$ is the transmitted symbol from $S$, $h_{ij,k}$ is the $ij$ link channel coefficient, $i\in(S,R_m)$ and $j\in(R_m,D)$, and $n_{ij,k}$ is the associated noise term. In this article, the destination is assumed to be affected by AWGN only, while the relays are subject to impulsive interference. This refers to the scenario where the sensor nodes acting as relays are located in the field of application generating the IN while the destination is the remote monitoring centre located in the far field. We assume that the channel coefficients of each $ij$ link follow a Rayleigh distribution and are static for one symbol duration, while they vary from one symbol to another. Therefore, $h_{ij,k}$ is modeled as a zero-mean, independent, circularly symmetric complex Gaussian (CSCG) random variable with variance $\Omega_{ij}\equiv\mathrm{E}\{|h_{ij}|^2\}=1/{\lambda_{ij}^\eta}$, where $\mathrm{E}\{\cdot\}$ denotes expectation operator, $\lambda_{ij}$ is the relative distance from $i$ to $j$, and $\eta$ is the path loss exponent \cite{laneman2004cooperative}. It is also assumed that the noise sample $n_{SR_m,k}$ follows the TSMG process that we will detail in the following subsection. We further assume that both the noise samples and the channel coefficients for each link are statistically independent. Unless otherwise explicitly mentioned, the instantaneous SNR of the $ij$ link is given by $\gamma_{ij}=P_i|h_{ij}|^2/\sigma_G^2$, where $\sigma_G^2$ represents the variance of the background Gaussian noise. The corresponding average SNR is given by $\overline{\gamma}_{ij}=P_i\Omega_{ij}/\sigma_G^2$.

In the second time slot, the $N$'th best relay $R_N$ demodulates the received signal $y_{SR_N}$ to recover the source information. Then, $R_N$ modulates the recovered signal using BPSK modulation and forwards it to the destination. The signal received at the destination node is therefore given by
\begin{equation}
y_{R_ND,k}=\sqrt{P_N}h_{R_ND,k} x_{R_N,k}+n_{{R_ND,}k},\label{signal_RbD_path}
\end{equation}
where $P_N$ is the average relay transmission power and $x_{R_N,k}$ is the forwarded symbol from $R_N$ which may be different from $x_{S,k}$ due to the possibility of decoding errors at the relay.
\subsection{Noise Model}
For a TSMG model, at each $k$, the statistical behavior of $n_{SR_m,k}$ is fully described by the noise state $s_{m,k} \in \{G,B\}$. In the context of our noise modeling, $G$ is referred to as the good state and $B$ as the bad state. The motivation of considering such a noise model stems from the fact that the good state happens when the channel is impaired by AWGN only, while the bad state takes place when this latter is subject to impulsive interference. For each $SR_m$ link, we model $n_{SR_m,k}$ as a zero-mean, independent, CSCG random variable, so that conditioned on $s_{m,k}$, the PDF of $n_{SR_m,k}$ can be expressed as
\begin{equation}
 f(n_{SR_m,k}|s_{m,k}\!=\!t)\!=\!\frac{1}{\pi \sigma_t^2} \exp\!\left(\!-\frac{|n_{SR_m,k}|^2}{\sigma_t^2}\!\right),\:t\in(G,B).\label{TSMG_model}
\end{equation}
Moreover, the parameter $\rho=\sigma_B^2/\sigma_G^2$ specifies the impulsive to Gaussian noise power ratio. The statistical description of the state process $\mathbf{s}_m^K\!=\!\{s_{m,0},s_{m,1},\ldots,s_{m,K-1}\}$ completely describes the channel and can be evaluated by the state transition probabilities $p_{s_{m,k}s_{m,k+1}}=p(s_{m,k+1}|s_{m,k})$, $s_{m,k},s_{m,k+1}\in\{G,B\}$. Given these transition probabilities, the stationary probability $p_G$ of being in the good state and $p_B$ of being in the bad state are respectively given by \cite{fertonani2009reliable},
\begin{equation}
p_G=\frac{p_{BG}}{p_{GB}+p_{BG}} \:\:\:\: \textrm{and} \:\:\:\: p_B=\frac{p_{GB}}{p_{GB}+p_{BG}}.
\end{equation}
It is worth mentioning that the parameter $\mu=\frac{1}{p_{GB}+p_{BG}}$ characterizes the noise memory and $\mu>1$ represents a channel that has a persistent memory.
\section{Relay Selection Protocols}\label{relay_techniques}
\subsection{Conventional Best Relay Selection Protocol}
As customary in the literature, for conventional BRS protocol, the best relay $R_b$ from the available $M$ relays is selected according to the following rule
\begin{equation}
R_b=\arg \max_{m\in\{1,2,\ldots,M\}}\left\{\min \left\{|h_{SR_m}|^2,|h_{R_mD}|^2\right\}\right\}.\label{bestrelay_AWGN}
\end{equation}
This \textit{max-min} BRS criterion establishes a tight upper bound in terms of end-to-end SNR \cite{bletsas2006simple}. Although this strategy exhibits the optimal performance for Gaussian environments, it may become inefficient in the presence of bursty IN since this \textit{max-min} BRS criterion relies on the channel statistics only and does not take into account the IN behavior when selecting the relay. Therefore, in the following section, we will propose a relay selection protocol for opportunistic relaying in the presence of bursty IN. The proposed protocol can be regarded as an extension of the conventional BRS protocol.
\subsection{Proposed Relay Selection Protocol in the Presence of Bursty Impulsive Noise}
\begin{figure}[!t]
  \centering
  \includegraphics[scale=0.53]{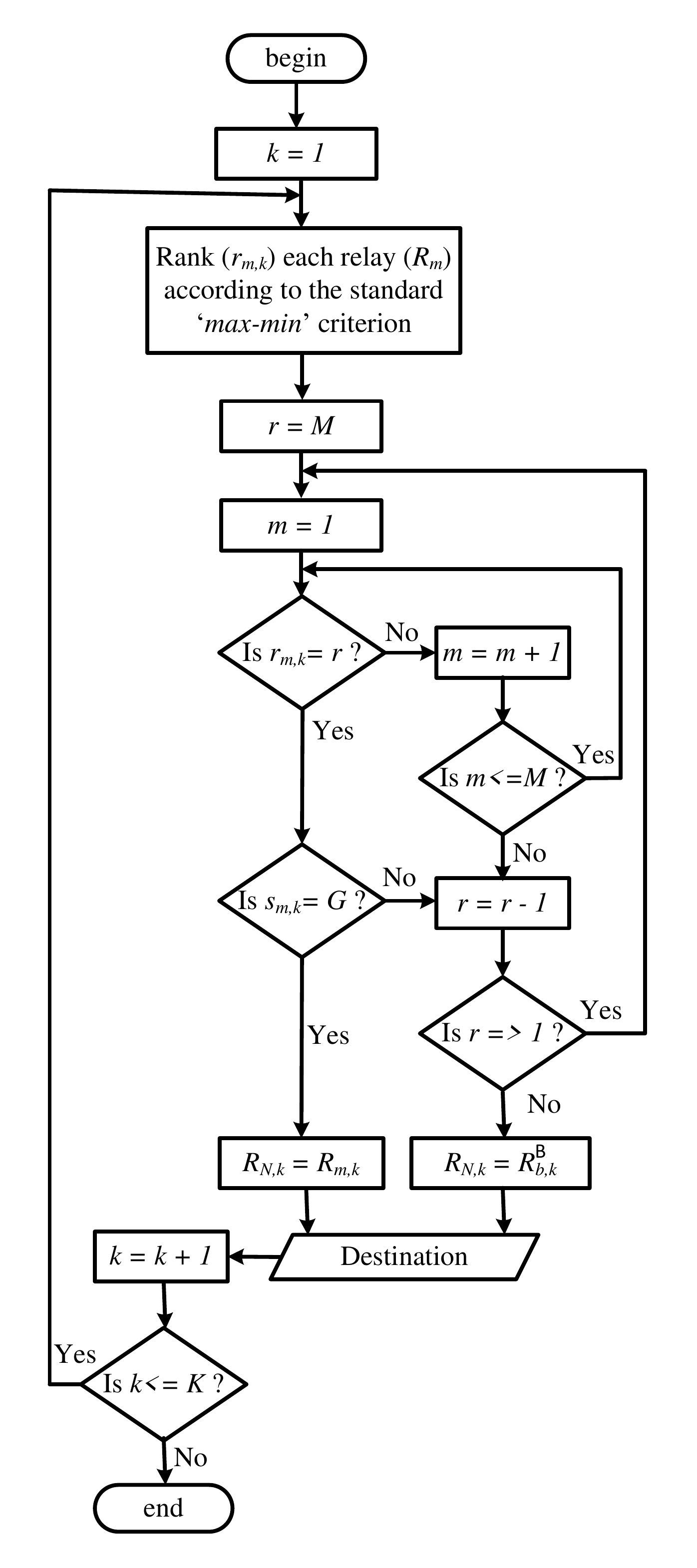}\\
  \caption{Flow diagram of the proposed $N$'th BRS protocol in the presence of bursty impulsive noise.}\label{flow_diagram_Mth_bestrelay}
\end{figure}
In this section, we focus on investigating the BRS in the presence of bursty IN. Since the conventional optimal BRS criterion cannot exploit the IN behavior, it may incur large performance degradation in the presence of strong interference at the relays. Hence, definite changes are required to the \textit{max-min} criterion to adapt to IN environments. On the other hand, if there is any way for each relay to know the state of the IN, the relay selection could be performed, based on the combined effect of the channel quality and the impulsive behavior. Given the IN state information, the conventional \textit{max-min} relay selection criterion can be extended to achieve the optimal performance. In this vein, from the implementation perspective, we assume that each relay has the ability to locally perform a noise state detection test at each time slot to determine whether it is affected by Gaussian noise or by impulsive state. When this state information is available at all the relays, a rational selection strategy would be as follows. First, rank ($r_{m,k}$) each relay ($R_m$) at time epoch $k$ according to the conventional \textit{max-min} criterion with the channel gain ordered in a non-increasing fashion. The relay $R_m$ in the first position of the ordered vector will be the best relay ($r_{m,k}\!=\!M$, full rank), the relay in the second position will be the second-best relay and so on. Then, the very next step is to check the state of the noise that affects the best relay. If the best relay is affected by impulsive state, try the second-best relay and so on. We termed this \textbf{$N$'}\textbf{th best relay selection} strategy for the proposed scenario. Finally, when all the relays are affected by impulsive state, choose the best relay that is in the impulsive state and has the bottleneck channel quality confirmed by (\ref{bestrelay_AWGN}). The received SNR at each relay under this condition becomes $\gamma_{SR_m}^B\!=\!\gamma_{SR_m}/\rho$. Hence, the conventional \textit{max-min} BRS criterion in (\ref{bestrelay_AWGN}) gives us
\begin{equation}
R_b^B=\arg\max_{m\in\{1,2,\ldots,M\}}\left\{\min \left\{\gamma_{SR_m}/\rho,\gamma_{R_mD}\right\}\right\}.\label{bestrelay_Imp}
\end{equation}
This new BRS criterion is very much dependent on the value of $\rho$ and for $\rho>>1$, it is highly likely that $\min \left\{\gamma_{SR_m}/\rho,\gamma_{R_mD}\right\}$ yields $\gamma_{SR_m}/\rho$. Thus, the BRS criterion in (\ref{bestrelay_Imp}) can be modified as
\begin{equation}
R_b^B|\rho>>1=\arg\max_{m\in\{1,2,\ldots,M\}}\left(\gamma_{SR_m}/\rho\right).\label{bestrelay_Imp_par}
\end{equation}
The selection criterion in (\ref{bestrelay_Imp_par}) is known as the partial BRS protocol \cite{krikidis2009max}. This is because this latter is dependent on the channel quality for the $SR$ link only and not on the end-to-end channel gains. It is shown in \cite{krikidis2009max} that this partial relay selection criterion poses the best performance from an asymptotic point of view.

The end-to-end steps of the proposed $N$'th BRS protocol are shown in Fig~\ref{flow_diagram_Mth_bestrelay}. As a consequence, in the following subsections, we detail different state detection algorithms to study the impact of the noise state information explicitly in the relay selection process.
\subsubsection{Genie detection}
Genie detection assumes that all the available relays have exact knowledge of the noise state. Although, this approach allows us to provide a tight limit of the best achievable performance, we observe that it is only conceptually valuable and the implementation of this detector is a very challenging task, if not impractical. In what follows, to reach the achievable performance, we propose some algorithms to obtain the states of IN.
\subsubsection{Proposed MAP based state detection algorithm}
To know the state of IN, in this scheme, at each $k$, each relay evaluates the a posteriori probability $p(s_{m,k}|y_{SR_m}^K)$ that the state $s_{m,k}$ is the actual channel state of relay $R_m$ at $k$, given the received sequence $y_{SR_m}^K=\{y_{SR_m,0},y_{SR_m,1},\ldots,y_{SR_m,K-1}\}$. This can be evaluated as
\begin{equation}
p\left(s_{m,k}|y_{SR_m}^K\right)\propto p\left(s_{m,k},y_{SR_m}^K\right).\label{aposteriori_state}
\end{equation}
Let us define the following quantities
\begin{equation}
\alpha_k(s_{m,k})=p\left(y_{SR_m,0},y_{SR_m,1},\ldots,y_{SR_m,k-1},s_{m,k}\right),\label{alpha}
\end{equation}
\begin{equation}
\beta_k(s_{m,k})=p\left(y_{SR_m,k},y_{SR_m,k+1},\ldots,y_{SR_m,K-1}|s_{m,k}\right),\label{beta}
\end{equation}
\begin{equation}
\delta_k(x_{S,k},s_{m,k},s_{m,k+1})=p\left(s_{m,k+1}|s_{m,k})p(n_{SR_m,k}|s_{m,k}\right),\label{gama}
\end{equation}
\begin{figure}[!t]
  \centering
  \includegraphics[scale=0.5]{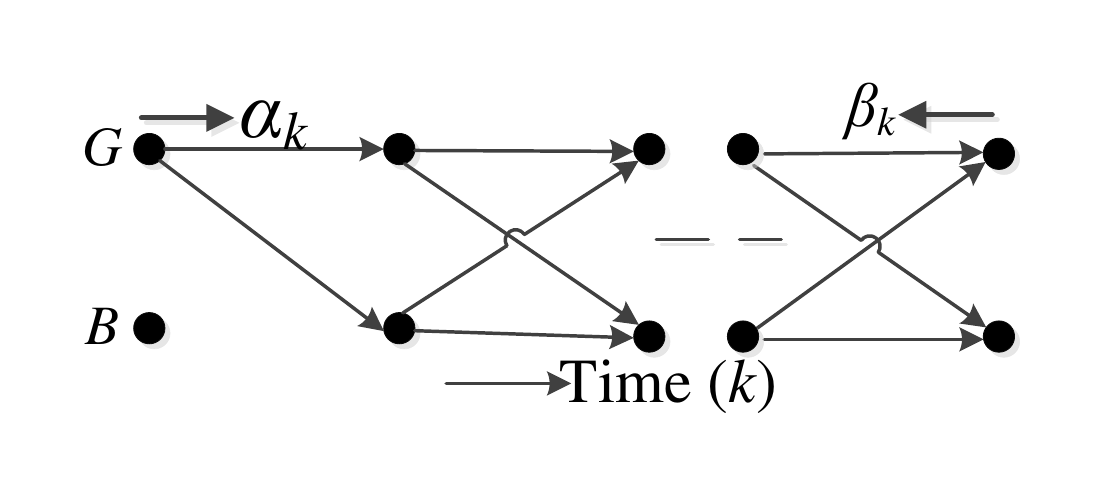}\\
  \caption{Trellis diagram for the representation of the TSMG noise model.}\label{trellisdiagram}
\end{figure}
where $\alpha_k(s_{m,k})$, $\beta_k(s_{m,k})$, and $\delta_k(x_{S,k},s_{m,k},s_{m,k+1})$ represent the forward filter, backward filter, and branch metrics of the trellis diagram, respectively, as shown in Fig.~\ref{trellisdiagram}. For the TSMG model, $p(n_{SR_m,k}|s_{m,k})$ in (\ref{gama}) can be evaluated using (\ref{TSMG_model}). Then, from (\ref{alpha}) and (\ref{beta}), $p(s_{m,k},y_{SR_m}^K)$ in (\ref{aposteriori_state}) can be expressed as
\begin{equation}
p(s_{m,k},y_{SR_m}^K)=\alpha_k(s_{m,k})\beta_k(s_{m,k}).
\end{equation}
Therefore, the state of the IN can be obtained as
\begin{eqnarray}
\hat{s}_{m,k}{~} = \left\{ \begin{array}{ll}
  G  & \textrm{if $L_{s_{m,k}} \geq 0$}\\
  B  & \textrm{if $L_{s_{m,k}} < 0$}
  \end{array} \right.\label{decision_rule}
\end{eqnarray}
where, $\hat{s}_{m,k}$ represents the estimate of $s_{m,k}$ (hard decision) and $L_{s_{m,k}}$ is the log-likelihood ratio (LLR). For this, the LLR values at the relays can be computed by
\begin{equation}
L_{s_{m,k}}=\ln \left\{ \frac{\alpha_k(s_{m,k}=G)\beta_k(s_{m,k}=G)}{\alpha_k(s_{m,k}=B)\beta_k(s_{m,k}=B)}\right\}.
\end{equation}
Accordingly, the algorithm computes the forward and backward filters recursively as
\begin{equation}
\alpha_{k+1}(s_{m,k+1})\!\!=\!\!\!\!\!\sum_{s_{m,k},x_{S,k}}\!\!\!\alpha_k(s_{m,k})p(x_{S,k})\delta_k(x_{S,k},s_{m,k},s_{m,k+1}),
\end{equation}
\begin{equation}
\beta_{k}(s_{m,k})\!\!=\!\!\!\!\!\!\!\sum_{s_{m,k+1},x_{S,k}}\!\!\!\!\!\!\beta_{k+1}(s_{m,k+1})p(x_{S,k})\delta_k(x_{S,k},s_{m,k},s_{m,k+1}),
\end{equation}
where we initialize the forward and backward filters as $\alpha_0(s_{m,0}=s)=p_s, \: \textrm{and}\: \beta_K(s_{m,K}=s)=1$, $s\in\{G,B\}$.
\subsubsection{Memoryless state detection}
Here, we consider a state detection algorithm known as memoryless state detection. Even though, this scheme is aware of the IN state, it cannot take into account the inherent noise memory. In this case, it is assumed that $\mu=1$ in the noise state detection process, which reflects the Bernoulli-Gaussian noise \cite{ghosh1996analysis} instead of TSMG noise. In this scenario, the previous MAP-based state detection algorithm is simplified to a sample-by-sample algorithm and the probability of being in a state will depend on $p(s_{m,k}|y_{SR_m,k})$, given by
\begin{equation}
p(s_{m,k}|y_{SR_m,k})\propto p(s_{m,k},y_{SR_m,k}),
\end{equation}
\begin{equation}
=p(s_{m,k})\sum_{x_{S,k}}p(n_{SR_m,k}|s_{m,k})p(x_{S,k}).
\end{equation}
Then, the LLR values at each relay can be obtained from
\begin{equation}
L_{s_{m,k}}=\ln \left\{\frac{p_G\sum_{x_{S,k}}p(n_{SR_m,k}|s_{m,k}=G)p(x_{S,k})}{p_B\sum_{x_{S,k}}p(n_{SR_m,k}|s_{m,k}=B)p(x_{S,k})}\right\}.
\end{equation}
From the LLR values, every relay then determines the noise states using (\ref{decision_rule}).

Although implementation related details are not our primary concern, in our scheme, the best relay can be selected either at the destination node in a centralized manner, or this selection can be performed distributively amongst the relays. For the first scheme, the channel state information (CSIs) of each SR and RD links, and the state of the impulsive noise of each SR links are required at the destination node. Similar to \cite{fareed2009relay}, it can be assumed that the destination node has the knowledge of $h_{SR_m}$ and $h_{R_mD}$ at the end of the first time slot. However, it is worth mentioning that the proposed scheme also needs to transmit the noise state information from the relays to the destination. The destination then prepares a ranking table of all the relays based on this information and chooses the best relay depending on the ranking information. Hence, an increase of signalling overhead is unavoidable for relay selection in impulsive environments.

For the implementation of a distributed scheme, similar to \cite{bletsas2006simple}, it is assumed that the relay nodes monitor the instantaneous channel conditions toward the source and the destination, and decide in a distributed fashion which one has the strongest path for information relaying. The best relay then checks its noise state information. When the best relay determines that it is in the impulsive state, it sends a beacon signal and the second best relay will check its impulsive state. The process continues until an interference free best relay is selected or to the point where all the relays are affected by impulsive state.
\subsection{Random Relay Selection Protocol}
In contrast to the previous relay selection protocols, for this protocol, one relay is picked randomly from all the available relays. This is suitable for simple scenario since its implementation neither requires the channel statistics nor the IN states and will probably show the worst performance.
\subsection{Complexity Discussion}
It is worth pointing out that, despite the performance increase, the complexity of the proposed MAP-based relay selection scheme grows exponentially with the frame length, due to the execution of the forward-backward algorithm, while it grows linearly in case of symbol-by-symbol selection schemes \cite{fertonani2007reduced}. For example, the complexity of MAP selection for an $M$-ary modulation system is $O(M^K)$, where $K$ is the frame length. On the other hand, the complexity of the symbol-by-symbol selection schemes are $O(K)$. However, in Section \ref{N_results} we show that the complexity of the MAP-based relay selection scheme is justified by its potential performance gain, making it a potential candidate for reliable communication scenarios. Hence, the proposed relay selection algorithm exhibits a performance/complexity trade-off.
\section{BER Performance Analysis}\label{ber_analysis}
In this section, we derive the BER expression of the proposed relay selection scheme under independent and identically distributed (i.i.d.) Rayleigh fading and bursty IN assuming that all the relays have perfect knowledge of IN state (genie-condition). We first consider the scenario where the selected relay is in good state. Since we assume that error can be propagated from the selected relay, the end-to-end error probability under this consideration can be expressed as
\begin{equation}
P_{e,D}(N)=P_{e,R_N}\cdot P_{e,SR_ND}^{er}+ (1-P_{e,R_N})\cdot P_{e,SR_ND}^{ner},\label{ber_atdestination}
\end{equation}
where $P_{e,R_N}$ is the error probability at the $N$'th best relay, $P_{e,SR_ND}^{er}$ is the destination error probability when an error is propagated from the $N$'th best relay, and $P_{e,SR_ND}^{ner}$ is the error probability at the destination when there is no error propagation from the $N$'th best relay.

Meanwhile, when all the relays are considered to be affected by IN, the system is forced to choose a best relay ($R_b^B$) that is in the  impulsive state and has the bottleneck channel quality confirmed by (\ref{bestrelay_Imp_par}). The overall BER performance will therefore be governed by the probability for which each of the selected relays transmits in either the good state or the bad state. For example, the first best relay will transmit in the good state with probability $(1-p_B)$ and the second best relay will transmit with probability $p_B(1-p_B)$ and so on. Finally, the probability of having all the available relays in bad state is $p_B^M$. The overall error probability at the destination is therefore given by
\begin{equation}
P_{e,D}=\sum_{N=1}^M (1-p_B)p_B^{N-1}P_{e,D}(N)+p_B^M P_{e,D}^B.\label{ber_etoe_des_final_GB}
\end{equation}
where $P_{e,D}^B$ is the destination error probability when all the relays are in bad state. As discussed, the first and second terms in (\ref{ber_etoe_des_final_GB}) represent respectively the overall probability of error at the destination when the selected relay is either in good state or in bad state.
\subsection{Calculation of $P_{e,D}(N)$}
\subsubsection{BER analysis at the $N$'th best relay}
The PDF of the received SNR from the source to the $N$'th best relay $\gamma_{SR_N}$ can be obtained by
\begin{eqnarray}
f_{\gamma_{SR_N}}(x)\!\!\!\!\!&=&\!\!\!\!\!\frac{C_M}{\overline{\gamma}_{R_mD}}\!\!\!\sum_{k=0}^{M-N}\!\!\!{{M-N}\choose{k}}(-1)^k \frac{\overline{\gamma}_{a}}{(k+N)\overline{\gamma}_{SR_m}\!\!-\!\overline{\gamma}_{a}}\nonumber\\
&&\times\left(e^{-x/\overline{\gamma}_{SR_m}}-e^{-x(k+N)/\overline{\gamma}_{a}}\right)\nonumber\\
&&\!\!+\frac{C_M}{\overline{\gamma}_{SR_m}}\!\!\!\!\sum_{k=0}^{M-N}\!\!\!{{M-N}\choose{k}}(-1)^k e^{-x(k+N)/\overline{\gamma}_{a}}, \label{marginal_PDF_sr_finite}
\end{eqnarray}
where $C_M=M{{M-1}\choose{N-1}}$ and $\overline{\gamma}_{a}=\frac{\overline{\gamma}_{SR_m}\overline{\gamma}_{R_mD}}{\overline{\gamma}_{SR_m}+\overline{\gamma}_{R_mD}}$.\\
\textit{Proof}:
The joint PDF of the $N$'th order statistics, $X_{(N)}=f_{\gamma_{SR_N}\gamma_{R_ND}}$ can be written as in (\ref{joint_pdf_12}) where $F(x)$ is the CDF of $f(x)$ \cite{papoulis2002probability,davidorder}.
\begin{figure*}
\begin{eqnarray}\label{joint_pdf_12}
f_{\gamma_{SR_N}\gamma_{R_ND}}(x,z) = \left\{ \begin{array}{ll}
   M{{M-1} \choose {N-1}}f_{\gamma_{SR_m}}(x)f_{\gamma_{R_mD}}(z)\left[F_{\gamma_{a}}(x)\right]^{M-N}\left[1-F_{\gamma_{a}}(x)\right]^{N-1}; & \textrm{if $x<z$}\\
  M{{M-1} \choose {N-1}}f_{\gamma_{SR_m}}(x)f_{\gamma_{R_mD}}(z)\left[F_{\gamma_{a}}(z)\right]^{M-N}\left[1-F_{\gamma_{a}}(z)\right]^{N-1}; & \textrm{if $x>z$}
  \end{array} \right.
\end{eqnarray}
\hrulefill
\end{figure*}
For Rayleigh fading channel, (\ref{joint_pdf_12}) can be rewritten as in (\ref{joint_distribution}).
\begin{figure*}
\begin{eqnarray}
f_{\gamma_{SR_N}\gamma_{R_ND}}(x,z) = \left\{ \begin{array}{ll}
   M{{M-1} \choose {N-1}}\frac{1}{\overline{\gamma}_{SR_m}}e^{-\frac{x}{\overline{\gamma}_{SR_m}}}\frac{1}{\overline{\gamma}_{R_mD}}e^{-\frac{z}{\overline{\gamma}_{R_mD}}}
\left[1-e^{-\frac{x}{\overline{\gamma}_{a}}}\right]^{M-N}\left[e^{-\frac{x}{\overline{\gamma}_{a}}}\right]^{N-1}; & \textrm{if $x<z$};\\
  M{{M-1} \choose {N-1}}\frac{1}{\overline{\gamma}_{SR_m}}e^{-\frac{x}{\overline{\gamma}_{SR_m}}}\frac{1}{\overline{\gamma}_{R_mD}}e^{-\frac{z}{\overline{\gamma}_{R_mD}}}
\left[1-e^{-\frac{z}{\overline{\gamma}_{a}}}\right]^{M-N}\left[e^{-\frac{z}{\overline{\gamma}_{a}}}\right]^{N-1}; & \textrm{if $x>z$}
  \end{array} \right.\label{joint_distribution}
\end{eqnarray}
\hrulefill
\end{figure*}
Now, from the joint distribution, the marginal distribution of $\gamma_{SR_N}$ can be obtained as
\begin{equation}
f_{\gamma_{SR_N}}(x)=\int_{z=0}^\infty f_{\gamma_{SR_N}\gamma_{R_ND}}(x,z)dz,\label{marginal_sr_main}
\end{equation}
Substituting (\ref{joint_distribution}) in (\ref{marginal_sr_main}), we get (\ref{exact_value}).
\begin{figure*}
\begin{eqnarray}
f_{\gamma_{SR_N}}(x)&=&\int_{z=0}^x M{{M-1} \choose {N-1}}\frac{1}{\overline{\gamma}_{SR_m}}e^{-\frac{x}{\overline{\gamma}_{SR_m}}}\frac{1}{\overline{\gamma}_{R_mD}}e^{-\frac{z}{\overline{\gamma}_{R_mD}}}
\left[1-e^{-\frac{z}{\overline{\gamma}_{a}}}\right]^{M-N}\left[e^{-\frac{z}{\overline{\gamma}_{a}}}\right]^{N-1} dz\nonumber\\
&+&\int_{z=x}^\infty M{{M-1} \choose {N-1}}\frac{1}{\overline{\gamma}_{SR_m}}e^{-\frac{x}{\overline{\gamma}_{SR_m}}}\frac{1}{\overline{\gamma}_{R_mD}}e^{-\frac{z}{\overline{\gamma}_{R_mD}}}
\left[1-e^{-\frac{x}{\overline{\gamma}_{a}}}\right]^{M-N}\left[e^{-\frac{x}{\overline{\gamma}_{a}}}\right]^{N-1}dz,\nonumber\\
&=& I_1 + I_2.\label{exact_value}
\end{eqnarray}
\hrulefill
\end{figure*}
Then, using the binomial expansion, $I_1$ in (\ref{exact_value}) can be written as
\begin{eqnarray}
I_1&=&\frac{M{{M-1} \choose {N-1}}}{\overline{\gamma}_{SR_m}\overline{\gamma}_{R_mD}}e^{-\frac{x}{\overline{\gamma}_{SR_m}}}\int_{z=0}^x e^{-\frac{z}{\overline{\gamma}_{R_mD}}}\sum_{k=0}^{M-N}{{M-N} \choose {k}}\nonumber\\
&&\times(-1)^k e^{-\frac{(k+N-1)z}{\overline{\gamma}_{a}}}dz,\label{I_1_finite}
\end{eqnarray}
Solving the integration and after some mathematical manipulations, (\ref{I_1_finite}) can be written as
\begin{eqnarray}
I_1\!\!\!&=&\!\!\!\frac{M{{M-1} \choose {N-1}}}{\overline{\gamma}_{R_mD}}\sum_{k=0}^{M-N}{{M-N} \choose {k}}(-1)^k \frac{\overline{\gamma}_{a}}{(k+N)\overline{\gamma}_{SR_m}\!-\!\overline{\gamma}_{a}}\nonumber\\
&&\times\left(e^{-x/\overline{\gamma}_{SR_m}}-e^{-x(k+N)/\overline{\gamma}_{a}}\right).\label{I_1_finite_final}
\end{eqnarray}
In a similar way, $I_2$ can be written as
\begin{eqnarray}
I_2\!\!\!\!&=&\!\!\!\!\frac{M{{M-1} \choose {N-1}}}{\overline{\gamma}_{SR_m}\overline{\gamma}_{R_mD}}e^{-\frac{x}{\overline{\gamma}_{SR_m}}}\!\!\!\sum_{k=0}^{M\!-\!N}\!\!\!{{M-N} \choose {k}}(-1)^k e^{-\frac{x(k+(N-1))}{\overline{\gamma}_{a}}}\nonumber\\
&&\times \int_{z=x}^\infty e^{-\frac{z}{\overline{\gamma}_{R_mD}}}dz,\nonumber\\
\!\!&=&\!\!\frac{M{{M-1}\choose {N-1}}}{\overline{\gamma}_{SR_m}}\sum_{k=0}^{M-N}{{M-N}\choose{k}}(-1)^k e^{-\frac{x(k+N)}{\overline{\gamma}_{a}}}.\label{I_2_finite_final}
\end{eqnarray}
Substituting the value of (\ref{I_1_finite_final}) and $(\ref{I_2_finite_final})$ in (\ref{exact_value}), (\ref{marginal_PDF_sr_finite}) is obtained. Therefore, the error probability of the source to the selected relay link can be obtained by \cite{proakis2001digital}
\begin{equation}
P_{e,R_N}=\frac{1}{2}\int_0^\infty \mathrm{erfc}(\sqrt{x})f_{\gamma_{SR_N}}(x)dx, \label{ber_sr_basic}
\end{equation}
where $f_{\gamma_{SR_N}}$ is provided in (\ref{marginal_PDF_sr_finite}) and $\mathrm{erfc}(\cdot)$ is the complementary error function. Solving the integral in (\ref{ber_sr_basic}) yields
\begin{eqnarray}
P_{e,R_N}\!\!\!&=&\!\!\!\frac{C_M}{\overline{\gamma}_{R_mD}}\!\!\sum_{k=0}^{M-N}\!\!{{M-N}\choose{k}}(-1)^k \frac{\overline{\gamma}_{a}}{(k+N)\overline{\gamma}_{SR_m}\!\!-\!\overline{\gamma}_{a}}\nonumber\\
&&\times \left[\omega\left(\frac{1}{\overline{\gamma}_{SR_m}}\right)-\omega\left(\frac{k+N}{\overline{\gamma}_{a}}\right)\right]\nonumber\\
&&+\frac{C_M}{\overline{\gamma}_{SR_m}}\!\!\sum_{k=0}^{M-N}\!\!{{M-N}\choose{k}}(-1)^k \omega\left(\frac{k+N}{\overline{\gamma}_{a}}\right).\label{ber_sr_finite}
\end{eqnarray}
To get (\ref{ber_sr_finite}), we use the identity $\omega(\theta)=\frac{1}{2\theta}\left[1-\frac{1}{\sqrt{1+\theta}}\right]$.
\subsubsection{BER analysis at the destination}
In order to compute $P_{e,SR_ND}^{er}$ and $P_{e,SR_ND}^{ner}$, we need the knowledge of the combining technique considered at the destination. For Gaussian channel, the maximum ratio combining (MRC) is optimal with regard to minimizing the BER \cite{proakis2001digital}. At this stage, since the selected $N$'th best relay is not affected by IN, we can perform MRC at the destination. The combined SNR at the destination, $\gamma_{SR_ND}$, is then the sum of two independent SNRs $\gamma_{SD}$ and $\gamma_{R_ND}$ with corresponding PDFs $f_{\gamma_{SD}}$ and $f_{\gamma_{R_ND}}$. Similar to (\ref{marginal_PDF_sr_finite}), the PDF of $\gamma_{R_ND}$ is given by
\begin{eqnarray}
f_{\gamma_{R_ND}}(x)\!\!\!\!\!&=&\!\!\!\!\!\frac{C_M}{\overline{\gamma}_{SR_m}}\!\!\!\sum_{k=0}^{M-N}\!\!\!{{M-N}\choose{k}}(-1)^k \frac{\overline{\gamma}_{a}}{(k+N)\overline{\gamma}_{R_mD}\!\!-\!\overline{\gamma}_{a}}\nonumber\\
&&\times\left(e^{-x/\overline{\gamma}_{R_mD}}-e^{-x(k+N)/\overline{\gamma}_{a}}\right)\nonumber\\
&&\!\!+\frac{C_M}{\overline{\gamma}_{R_mD}}\!\!\!\!\sum_{k=0}^{M-N}\!\!\!{{M-N}\choose{k}}(-1)^k e^{-x(k+N)/\overline{\gamma}_{a}}, \label{marginal_PDF_rd_finite}
\end{eqnarray}
Also, the PDF of $f_{\gamma_{SD}}$ is
\begin{equation}
f_{\gamma_{SD}}(y)=\frac{1}{\overline{\gamma}_{SD}}e^{-y/\overline{\gamma}_{SD}},
\end{equation}
Therefore, the PDF of $\gamma_{SR_ND}=\gamma_{SD}+\gamma_{R_ND}$ can be obtained by the well-known convolution theorem as
\begin{equation}
f_{\gamma_{SR_ND}}(\theta)=\int_0^\theta f_{\gamma_{R_ND}}(z)f_{\gamma_{SD}}(\theta-z)dz\label{convolution_SRD},
\end{equation}
which is expressed in (\ref{snr_srd_finite}).
\begin{figure*}
\begin{eqnarray}\label{snr_srd_finite}
f_{\gamma_{SR_ND}}(\theta)&=&\frac{C_M}{\overline{\gamma}_{SR_m}}\sum_{k=0}^{M-N}{{M-N} \choose {k}}(-1)^k \frac{\overline{\gamma}_{a}}{(k+N)\overline{\gamma}_{R_mD}-\overline{\gamma}_{a}}\nonumber\\
&&\times\left[\frac{\overline{\gamma}_{R_mD}}{\overline{\gamma}_{SD}-\overline{\gamma}_{R_mD}}\left(e^{-\theta/\overline{\gamma}_{SD}}-e^{-\theta/\overline{\gamma}_{R_mD}}\right)-\frac{\overline{\gamma}_{a}}{(k+N)\overline{\gamma}_{SD}-\overline{\gamma}_{a}}\left(e^{-\theta/\overline{\gamma}_{SD}}-e^{-(k+N)\theta/\overline{\gamma}_{a}}\right)\right]\nonumber\\
&&+\frac{C_M}{\overline{\gamma}_{R_mD}}\sum_{k=0}^{M-N}{{M-N} \choose {k}}(-1)^k \frac{\overline{\gamma}_{a}}{(k+N)\overline{\gamma}_{SD}-\overline{\gamma}_{a}}\left(e^{-\theta/\overline{\gamma}_{SD}}-e^{-(k+N)\theta/\overline{\gamma}_{a}}\right).
\end{eqnarray}
\hrulefill
\end{figure*}
Then, the error probability of the combined path assuming there is no error propagated from the selected relay is obtained by
\begin{equation}
P_{e,SR_ND}^{ner}=\frac{1}{2}\int_0^\infty \mathrm{erfc}(\sqrt{\theta})f_{\gamma_{SR_ND}}(\theta)d\theta,\label{ber_basic_eqn}
\end{equation}
which is shown in (\ref{ber_mrc_finite}).
\begin{figure*}
\begin{eqnarray}
P_{e,SR_ND}^{ner}&=&\frac{C_M}{\overline{\gamma}_{SR_m}}\sum_{k=0}^{M-N}{{M-N} \choose {k}}(-1)^k \frac{\overline{\gamma}_{a}}{(k+N)\overline{\gamma}_{R_mD}-\overline{\gamma}_{a}}\nonumber\\
&&\times\left[\frac{\overline{\gamma}_{R_mD}}{\overline{\gamma}_{SD}-\overline{\gamma}_{R_mD}}\left(\omega\left(\frac{1}{\overline{\gamma}_{SD}}\right)-\omega\left(\frac{1}{\overline{\gamma}_{R_mD}}\right)\right)-\frac{\overline{\gamma}_{a}}{(k+N)\overline{\gamma}_{SD}-\overline{\gamma}_{a}}\left(\omega\left(\frac{1}{\overline{\gamma}_{SD}}\right)-\omega\left(\frac{k+N}{\overline{\gamma}_{a}}\right)\right)\right]\nonumber\\
&&+\frac{C_M}{\overline{\gamma}_{R_mD}}\sum_{k=0}^{M-N}{{M-N} \choose {k}}(-1)^k \frac{\overline{\gamma}_{a}}{(k+N)\overline{\gamma}_{SD}-\overline{\gamma}_{a}}\left(\omega\left(\frac{1}{\overline{\gamma}_{SD}}\right)-\omega\left(\frac{k+N}{\overline{\gamma}_{a}}\right)\right).\label{ber_mrc_finite}
\end{eqnarray}
\hrulefill
\end{figure*}
From (\ref{ber_atdestination}), it is seen that we also need the expression of $P_{e,SR_ND}^{er}$, which can be tightly approximated for the considered BPSK modulated system as \cite{tourki2013new}
\begin{equation}
P_{e,SR_ND}^{er}\approx \frac{\overline{\gamma}_{R_ND}}{\overline{\gamma}_{R_ND}+\overline{\gamma}_{SD}},\label{error_des_witherror}
\end{equation}
where $\overline{\gamma}_{R_ND}$ is the expected value of $\gamma_{R_ND}$ and is given by
\begin{equation}
\overline{\gamma}_{R_ND}=\int_0^\infty \gamma_{R_ND}(z)f_{\gamma_{R_ND}}(z)dz,\label{avgsnr_rNd_link}
\end{equation}
So, from (\ref{avgsnr_rNd_link}) and (\ref{marginal_PDF_rd_finite}) $\overline{\gamma}_{R_ND}$ is obtained in (\ref{avg_rd_finite}).
\begin{figure*}
\begin{eqnarray}
\overline{\gamma}_{R_ND}&=&\frac{C_M}{\overline{\gamma}_{SR_m}}\sum_{k=0}^{M-N}{{M-N} \choose {k}}(-1)^k \frac{\overline{\gamma}_{a}}{(k+N)\overline{\gamma}_{R_mD}-\overline{\gamma}_{a}}\left[\overline{\gamma}_{R_mD}^2-\left(\frac{\overline{\gamma}_{a}}{k+N}\right)^2\right]\nonumber\\
&+&\frac{C_M}{\overline{\gamma}_{R_mD}}\sum_{k=0}^{M-N}{{M-N} \choose {k}}(-1)^k \left(\frac{\overline{\gamma}_{a}}{k+N}\right)^2.\label{avg_rd_finite}
\end{eqnarray}
\hrulefill
\end{figure*}
Therefore, the end-to-end error probability under the $N$'th BRS strategy when the selected relay is in good state can be evaluated by substituting (\ref{ber_sr_finite}), (\ref{ber_mrc_finite}), and (\ref{error_des_witherror}) in (\ref{ber_atdestination}).
\subsection{Calculation of $P_{e,D}^B$}\label{error_destination_imp}
Similar to (\ref{ber_atdestination}), the end-to-end probability of error when the selected relay is in bad state can be expressed as
\begin{equation}
P_{e,D}^B=P_{e,R_b^B}\cdot P_{e,SR_b^BD}^{er}+ (1-P_{e,R_b^B})\cdot P_{e,SR_b^BD}^{ner},\label{ber_atdestination_imp}
\end{equation}
Now, the PDF of the received SNR from the source to the best relay $\gamma_{SR_b^B}^B$ under this condition can be expressed as \cite{papoulis2002probability}
\begin{eqnarray}
f_{\gamma_{SR_b^B}^B}(y)&\!\!\!=\!\!\!&M F_x^{M-1}(y)f_x(y),\nonumber\\
&\!\!\!=\!\!\!& M\!\!\left(\!1-e^{-y/\overline{\gamma}_{SR_m}^B}\!\!\right)^{M-1}\!\!\frac{1}{\overline{\gamma}_{SR_m}^B}e^{-y/\overline{\gamma}_{SR_m}^B},\label{gamma_SR_b_finite}
\end{eqnarray}
Using the Binomial expansion, (\ref{gamma_SR_b_finite}) can be expressed as
\begin{equation}
f_{\gamma_{SR_b^B}^B}(y)=\frac{M}{\overline{\gamma}_{SR_m}^B}\sum_{k=0}^{M-1}{{M-1}\choose{k}}(-1)^k e^{-ky/\overline{\gamma}_{SR_m}^B}.\label{pdf_sr_imp_finite}
\end{equation}
Therefore, the error probability at the selected relay can be obtained as
\begin{equation}
P_{e,R_b^B}=\frac{M}{\overline{\gamma}_{SR_m}^B}\sum_{k=0}^{M-1}{{M-1}\choose{k}}(-1)^k\omega\left(\frac{k}{\overline{\gamma}_{SR_m}^B}\right).\label{ber_sr_imp_finite}
\end{equation}
Now, the BER at the destination can be obtained according to (\ref{ber_atdestination_imp}). It is assumed that the combining at the destination is based on MRC. Hence, $P_{e,SR_b^BD}^{ner}$ is the BER of a two-branch MRC receiver. For i.i.d. Rayleigh channels, this is given as \cite{proakis2001digital}
\begin{equation}
P_{e,SR_b^BD}^{ner}=\frac{1}{2}\left(\frac{\tau(\bar{\gamma}_{SD})}{1-\bar{\gamma}_{R_b^BD}/\bar{\gamma}_{SD}}+\frac{\tau(\bar{\gamma}_{R_b^BD})}{1-\bar{\gamma}_{SD}/\bar{\gamma}_{R_b^BD}}\right),\label{error_MRC_des_impulse}
\end{equation}
where $\tau(\bar{\gamma})=1-\sqrt{\frac{\bar{\gamma}}{1+\bar{\gamma}}}$ and $\bar{\gamma}_{R_b^BD}=\bar{\gamma}_{R_mD}$, since the second phase is independent of the relay selection process. In addition, similar to (\ref{error_des_witherror}), the error probability $P_{e,SR_b^BD}^{er}$, under this condition can be approximated by
\begin{equation}
P_{e,SR_b^BD}^{er}\approx \frac{\overline{\gamma}_{R_mD}}{\overline{\gamma}_{R_mD}+\overline{\gamma}_{SD}}.\label{error_prone_des_impulse}
\end{equation}
Finally, substituting (\ref{ber_sr_imp_finite}), (\ref{error_MRC_des_impulse}), and (\ref{error_prone_des_impulse}) in (\ref{ber_atdestination_imp}) $P_{e,D}^B$ can be obtained.
\section{Outage analysis}\label{outage_analysis}
The end-to-end outage probability of the proposed scheme for a data rate $R$ when the selected relay in good state is given by \cite{tourki2013new}
\begin{eqnarray}
P_{out}(N)\!\!\!\!&=&\!\!\!\!p\left\{\gamma_{SR_N}\!>\!\phi,\gamma_{R_ND}+\gamma_{SD}\!<\!\phi\right\}\!+\!p\left\{\gamma_{SR_N}\!<\!\phi\right\}\nonumber\\
&&\times p\left\{\gamma_{SD}<\phi\right\},\label{prob_outage_Nbest}
\end{eqnarray}
where $\phi=2^{2R}-1$. Therefore, the overall outage probability at the destination is given by
\begin{equation}
P_{out}=\sum_{N=1}^M (1-p_B)p_B^{N-1}P_{out}(N)+p_B^M P_{out}^B.\label{p_out_final}
\end{equation}
where $P_{out}^B$ is the outage probability at the destination when all the $M$ relays are in bad state and therefore the selected relay is in bad state as well.
\subsection{Calculation of $P_{out}(N)$}\label{outage_des_good_finite}
Now, the outage probability at the $N$'th best relay is obtained by
\begin{eqnarray}
P_{out,SR_N}\!\!\!\!&=&\!\!\!\!\int_0^\phi f_{\gamma_{SR_N}}(x)dx \equiv F_{\gamma_{SR_N}}(\phi),\nonumber\\
\!\!\!&=&\!\!\!\frac{C_M}{\overline{\gamma}_{R_mD}}\!\!\sum_{k=0}^{M-N}\!\!{{M-N}\choose{k}}(-1)^k \frac{\overline{\gamma}_{a}}{(k+N)\overline{\gamma}_{SR_m}\!\!-\!\overline{\gamma}_{a}}\nonumber\\
&&\times \left[\chi_{\overline{\gamma}_{SR_m}}\left(\phi\right)-\chi_{\frac{\overline{\gamma}_{a}}{k+N}}\left(\phi\right)\right]\nonumber\\
&&+\frac{C_M}{\overline{\gamma}_{SR_m}}\!\!\sum_{k=0}^{M-N}\!\!{{M-N}\choose{k}}(-1)^k \chi_\frac{\overline{\gamma}_{a}}{k+N}\left(\phi\right),\label{p_out_SRN_finite}
\end{eqnarray}
where $F_{\gamma_{SR_N}}(x)$ is the CDF of $\gamma_{SR_N}(x)$ shown in (\ref{marginal_PDF_sr_finite}) and $\chi_a(x)=a\left(1-e^{-x/a}\right)$. Similarly, the outage probability for the $SD$ link becomes
\begin{eqnarray}
P_{out,SD}=F_{\gamma_{SD}}(\phi)=\frac{\chi_{\overline{\gamma}_{SD}}\left(\phi\right)}{\overline{\gamma}_{SD}},\label{p_out_SD_finite}
\end{eqnarray}
On the other hand, the first term in (\ref{prob_outage_Nbest}) can be approximated as \cite{tourki2013new}
\begin{equation}
p\left\{\gamma_{SR_N}\!>\!\phi,\gamma_{R_ND}\!+\!\gamma_{SD}\!<\!\phi\right\}\!\approx\!\left(\!1\!-\!F_{\gamma_{SR_N}}(\phi)\!\right)\!\!F_{\gamma_{SR_ND}}(\phi),
\end{equation}
where $F_{\gamma_{SR_N}}(\phi)$ can be derived according to (\ref{p_out_SRN_finite}). Also, the outage probability for the $SR_ND$ link can be obtained by taking the CDF of (\ref{snr_srd_finite}) as shown in (\ref{outage_srd_finite}). Hence, the end-to-end outage probability when the selected relay is in good state can be evaluated by substituting (\ref{p_out_SRN_finite}), (\ref{p_out_SD_finite}), and (\ref{outage_srd_finite}) in (\ref{prob_outage_Nbest}).
\begin{figure*}
\begin{eqnarray}\label{outage_srd_finite}
P_{out,SR_ND}&=&\frac{C_M}{\overline{\gamma}_{SR_m}}\sum_{k=0}^{M-N}{{M-N} \choose {k}}(-1)^k \frac{\overline{\gamma}_{a}}{(k+N)\overline{\gamma}_{R_mD}-\overline{\gamma}_{a}}\nonumber\\
&&\times\left[\frac{\overline{\gamma}_{R_mD}}{\overline{\gamma}_{SD}-\overline{\gamma}_{R_mD}}\left(\chi_{\overline{\gamma}_{SD}}\left(\phi\right)-\chi_{\overline{\gamma}_{R_mD}}\left(\phi\right)\right)-\frac{\overline{\gamma}_{a}}{(k+N)\overline{\gamma}_{SD}-\overline{\gamma}_{a}}\left(\chi_{\overline{\gamma}_{SD}}\left(\phi\right)-\chi_{\frac{\overline{\gamma}_{a}}{k+N}}\left(\phi\right)\right)\right]\nonumber\\
&&+\frac{C_M}{\overline{\gamma}_{R_mD}}\sum_{k=0}^{M-N}{{M-N} \choose {k}}(-1)^k \frac{\overline{\gamma}_{a}}{(k+N)\overline{\gamma}_{SD}-\overline{\gamma}_{a}}\left(\chi_{\overline{\gamma}_{SD}}\left(\phi\right)-\chi_{\frac{\overline{\gamma}_{a}}{k+N}}\left(\phi\right)\right).
\end{eqnarray}
\hrulefill
\end{figure*}
\subsection{Calculation of $P_{out}^B$}\label{outage_des_bad_finite}
Similar to (\ref{prob_outage_Nbest}), the end-to-end outage probability when the selected relay is in bad state can be obtained by
\begin{equation}
P_{out}^B=P_{out,SR_b^B}\cdot P_{out,SD}+ (1-P_{out,SR_b^B})\cdot P_{out,SR_b^BD},\label{outage_atdestination_imp}
\end{equation}
where $P_{out,SR_b^B}$ can be obtained by taking the CDF of (\ref{pdf_sr_imp_finite}) yielding
\begin{equation}
P_{out,SR_b^B}=\frac{M}{\overline{\gamma}_{SR_m}^B}\sum_{k=0}^{M-1}{{M-1}\choose{k}}(-1)^k \chi_{\overline{\gamma}_{SR_m}^B}\left(k\phi\right).\label{outage_sr_imp_finite}
\end{equation}
Moreover, the outage probability $P_{out,SR_b^BD}$ at the destination under this condition can be obtained as \cite{goldsmith2005wireless}
\begin{equation}
P_{out,SR_b^BD}=\frac{1}{2}\frac{\chi_{\overline{\gamma}_{SD}}\left(\phi\right)\chi_{\overline{\gamma}_{R_mD}}\left(\phi\right)}{\overline{\gamma}_{SD}\overline{\gamma}_{R_mD}}.\label{outage_MRC_des_impulse}
\end{equation}
Finally, substituting (\ref{outage_sr_imp_finite}), (\ref{outage_MRC_des_impulse}), and (\ref{p_out_SD_finite}) in (\ref{outage_atdestination_imp}), $P_{out}^B$ can be evaluated.
\section{Asymptotic analysis}\label{asymptotic_analysis}
To provide more insights on the system behavior, we here reformulate the asymptotic BER and outage analysis for the proposed relay selection scheme. This allows us to validate the simulation results in high SNR regions.
\subsection{Asymptotic BER analysis}
\subsubsection{Asymptotic equivalence of $P_{e,R_N}$}
We show in Appendix A that the asymptotic PDF of $\gamma_{SR_N}$ can be expressed as
\begin{equation}
f_{\gamma_{SR_N}}(x)\doteq M{{M-1} \choose {N-1}}\frac{1}{\overline{\gamma}_{SR_m}}\left(\frac{1}{\overline{\gamma}_{a}}\right)^{M-N} x^{M-N},\label{marginal_PDF_sr}
\end{equation}
where $\doteq $ denotes the asymptotic equality. Then, the probability of error for the source to the selected relay link can be derived according to (\ref{ber_sr_basic}) and becomes
\begin{equation}
P_{e,R_N}\doteq M{{M-1} \choose {N-1}}\frac{1}{\overline{\gamma}_{SR_m}}\left(\frac{1}{\overline{\gamma}_{a}}\right)^{M-N}\!\!\frac{\Gamma(M-N+3/2)}{2\sqrt{\pi}(M-N+1)},\label{ber_sr_final}
\end{equation}
where $\Gamma(\cdot)$ is the complete Gamma function. To get the closed-form expression in (\ref{ber_sr_final}), we use the following identities
\begin{equation}
\mathrm{erfc}(z)=\frac{\Gamma(1/2,z^2)}{\sqrt{\pi}}, \:\:\:\textrm{and}\:\int_0^\infty x^{a-1}\Gamma(b,x)dx=\frac{\Gamma(a+b)}{a}.
\end{equation}
\subsubsection{Asymptotic equivalence of $P_{e,SR_ND}^{ner}$ and $P_{e,SR_ND}^{er}$}
To obtain the asymptotic end-to-end BER at the destination according to (\ref{ber_atdestination}), we also need the asymptotic equivalence of $P_{e,SR_ND}^{er}$ and $P_{e,SR_ND}^{ner}$ which further requires the asymptotic PDF of $\gamma_{R_ND}$ and $\gamma_{SR_ND}$. Similar to (\ref{marginal_PDF_sr}), the PDF of $\gamma_{R_ND}$ is given by
\begin{equation}
f_{\gamma_{R_ND}}(z)\doteq M{{M-1} \choose {N-1}}\frac{1}{\overline{\gamma}_{R_mD}}\left(\frac{1}{\overline{\gamma}_{a}}\right)^{M-N} z^{M-N}.\label{marginal_PDF_rd}
\end{equation}
Therefore, the PDF of $\gamma_{SR_ND}$ can be obtained according to the convolution theorem depicted in (\ref{convolution_SRD}) as
\begin{eqnarray}
f_{\gamma_{SR_ND}}(\theta)&\doteq&\!\!\!\!M{{M-1} \choose {N-1}}\frac{1}{\overline{\gamma}_{R_mD}}\left(\frac{1}{\overline{\gamma}_{a}}\right)^{M-N}\frac{1}{\overline{\gamma}_{SD}}e^{-\theta/\overline{\gamma}_{SD}}\nonumber\\
&&\times \int_0^\theta z^{M-N} e^{z/\overline{\gamma}_{SD}}dz,\label{conv_gamma_sd_rd}
\end{eqnarray}
Integrating by parts and following some mathematical manipulations, (\ref{conv_gamma_sd_rd}) can be approximated as
\begin{equation}
f_{\gamma_{SR_ND}}(\theta)\approx \frac{M{{M-1} \choose {N-1}}}{(M-N+1)} \frac{1}{\overline{\gamma}_{R_mD}}\left(\frac{1}{\overline{\gamma}_{a}}\right)^{M-N}\frac{\theta^{M-N+1}}{\overline{\gamma}_{SD}}. \label{joint_distribution_SRD}
\end{equation}
Then, the error probability of the combined path when there is no error propagation from the selected relay is obtained by
\begin{eqnarray}
P_{e,SR_ND}^{ner}&=&\frac{1}{2}\int_0^\infty \mathrm{erfc}(\sqrt{\theta})f_{\gamma_{SR_ND}}(\theta)d\theta,\label{ber_basic_eqn}\\
&\doteq&\frac{M{{M-1} \choose {N-1}}}{2\sqrt{\pi}(M-N+1)}\frac{1}{\overline{\gamma}_{R_mD}}\left(\frac{1}{\overline{\gamma}_{a}}\right)^{M-N}\frac{1}{\overline{\gamma}_{SD}}\nonumber\\
&&\times\frac{\Gamma(M-N+5/2)}{M-N+2}.\label{error_des_MRC}
\end{eqnarray}
On the other hand, $P_{e,SR_ND}^{er}$ can be derived according to (\ref{error_des_witherror}), where $\overline{\gamma}_{R_ND}$ can be obtained as
\begin{equation}
\overline{\gamma}_{R_ND}=\int_0^\infty \gamma_{R_ND}(z)f_{\gamma_{R_ND}}(z)dz,\label{avgsnr_rMd_link}
\end{equation}
So, from (\ref{avgsnr_rMd_link}) and (\ref{marginal_PDF_rd}), we have
\begin{eqnarray}
\overline{\gamma}_{R_ND}\!\!\!\!\!&\doteq&\!\!\!\!\!M{{M-1} \choose {N-1}}\frac{1}{\overline{\gamma}_{R_mD}}\left(\frac{1}{\overline{\gamma}_{a}}\right)^{M-N}\!\!\!\!\!\int_0^\infty \!\!\!\!\!z^{M-N+1}e^{-z/\overline{\gamma}_{a}}dz,\nonumber\\
&=& \!\!\!\!M{{M-1} \choose {N-1}}\frac{1}{\overline{\gamma}_{R_mD}}\overline{\gamma}_{a}^2 \Gamma(M-N+2).
\end{eqnarray}
Therefore, the asymptotic end-to-end error probability under the assumption that the $N$'th best relay is in good state can be evaluated by substituting (\ref{ber_sr_final}), (\ref{error_des_MRC}), and (\ref{error_des_witherror}) in (\ref{ber_atdestination}).
\subsubsection{Asymptotic equivalence of $P_{e,D}^B$}\label{error_destination_impulsive}
From (\ref{gamma_SR_b_finite}), the asymptotic PDF of $\gamma_{SR_b^B}^B$ can be expressed as
\begin{equation}
f_{\gamma_{SR_b^B}^B}(y)\doteq M \left(\frac{1}{\overline{\gamma}_{SR_m}^B}\right)^M y^{M-1}.\label{gamma_SR_b_asymp}
\end{equation}
Therefore, the probability of error at the selected relay under this condition can be obtained by
\begin{equation}
P_{e,SR_b^B}\doteq \frac{1}{2\sqrt{\pi}}\left(\frac{1}{\overline{\gamma}_{SR_m}^B}\right)^M \Gamma(M+1/2).\label{error_bestrelay_impulse}
\end{equation}
On the other hand, the values of $P_{e,SR_b^BD}^{ner}$ and $P_{e,SR_b^BD}^{er}$ can be obtained according to (\ref{error_MRC_des_impulse}) and (\ref{error_prone_des_impulse}), respectively. Finally, substituting the value of (\ref{error_bestrelay_impulse}), (\ref{error_MRC_des_impulse}), and (\ref{error_prone_des_impulse}) in (\ref{ber_atdestination_imp}) the asymptotic expression of $P_{e,D}^B$ can be obtained.
\subsection{Asymptotic Outage Analysis}
\subsubsection{Asymptotic equivalence of $P_{out}(N)$}\label{outage_des_good_asymp}
The asymptotic outage probability at the $N$'th best relay is obtained by
\begin{eqnarray}
P_{out,SR_N}\!\!\!\!&=&\!\!\!\!\int_0^\phi f_{\gamma_{SR_N}}(x)dx \equiv F_{\gamma_{SR_N}}(\phi),\nonumber\\
&\doteq& \!\!\!\!M{{M-1} \choose {N-1}}\!\frac{1}{\overline{\gamma}_{SR_m}}\!\left(\!\frac{1}{\overline{\gamma}_{a}}\!\right)^{M-N}\!\!\!\!\frac{\phi^{M-N+1}}{M\!-\!N\!+\!1}.\label{p_out_SRN}
\end{eqnarray}
Similarly, the outage probability for the $SD$ link becomes
\begin{equation}
P_{out,SD}=F_{\gamma_{SD}}(\phi)\doteq \frac{\phi}{\overline{\gamma}_{SD}}.
\label{p_out_SD_asymp}
\end{equation}
Also, the outage probability for the $SR_ND$ link can be obtained from the joint distribution derived in (\ref{joint_distribution_SRD}) as
\begin{equation}
P_{out,SR_ND}\doteq\frac{M{{M-1} \choose {N-1}}}{M\!-\!N\!+\!1}\frac{1}{\overline{\gamma}_{R_mD}}\left(\!\frac{1}{\overline{\gamma}_{a}}\!\right)^{M-N}\!\!\frac{1}{\overline{\gamma}_{SD}}\frac{\phi^{M-N+2}}{M\!-\!N\!+\!2}.\label{p_out_SRND}
\end{equation}
Hence, the end-to-end outage probability can be evaluated by substituting (\ref{p_out_SRN}), (\ref{p_out_SD_asymp}), and (\ref{p_out_SRND}) in (\ref{prob_outage_Nbest}) and becomes
\begin{eqnarray}
P_{out}(N)&\doteq& \frac{M{{M-1} \choose {N-1}}}{M-N+1}\frac{1}{\overline{\gamma}_{SD}}\left(\frac{1}{\overline{\gamma}_{a}}\right)^{M-N}\phi^{M-N+2}\nonumber\\
&&\times\left[\frac{1}{M-N+2}\frac{1}{\overline{\gamma}_{R_mD}}+ \frac{1}{\overline{\gamma}_{SR_m}}\right].\label{outage_asymp_final}
\end{eqnarray}
\subsubsection{Asymptotic equivalence of $P_{out}^B$}\label{outage_des_bad_asymp}
The asymptotic equivalence of $P_{out}^B$ can be obtained according to (\ref{outage_atdestination_imp}), where the outage probability at the selected relay can be achieved by taking the CDF of (\ref{gamma_SR_b_asymp}) and is equal to
\begin{equation}
P_{out,SR_b^B}\doteq \left(\frac{\phi}{\overline{\gamma}_{SR_m}^B}\right)^M.
\end{equation}
Moreover, the outage probability $P_{out,SR_b^BD}$ at the destination under this condition can be approximated as \cite{goldsmith2005wireless}
\begin{equation}
P_{out,SR_b^BD}\doteq\frac{1}{\overline{\gamma}_{SD}}\frac{1}{\overline{\gamma}_{R_mD}}\left(\frac{\phi^2}{2}\right).
\end{equation}
From (\ref{outage_asymp_final}), we observe that the maximum achievable diversity order converges to $M-N+2$. Hence, the proposed $N$'th BRS scheme will achieve the full diversity order of $M+1$ when $N=1$, i.e., the proposed protocol chooses the first best relay for cooperation.
\section{Numerical results}\label{N_results}
In this section, we simulate the BER and the outage performances of the proposed DF relay selection schemes to validate the theoretical results presented in Section \ref{ber_analysis}, \ref{outage_analysis}, and \ref{asymptotic_analysis}. In our simulations, it is assumed that a frame of $10,000$ bits is mapped to a BPSK modulation sequence. It is then transmitted over Rayleigh quasi-static flat fading channels where the received sequence at the relays are impaired by TSMG noise characterized by $p_B=0.01$, $\mu=100$, and $\rho=100$ for each link. In this model, the $N$'th BRS is performed among a total number of $M=5$ relays and equal transmission power is considered at both the source and the selected relay. Moreover, we assume that $\lambda_{SD}=1$ and $\lambda_{SR_m}=0.4,\forall m$, where the relays are uniformly distributed between the $SD$ pair. The BER and the outage performances are calculated as a function of $SNR$ which is defined as, $SNR=E\{|x_{S,k}|^2|h_{ij}|^2\}/\sigma_G^2$. Furthermore, we assume that the the noise parameters ($p_B,\mu,\rho,\sigma_G^2$) and the channel coefficients $h_{ij}$ are perfectly known at the receiver. Finally, we set the path loss exponent to $\eta=2$.
\begin{figure}[!t]
  \centering
  \includegraphics[width=\columnwidth]{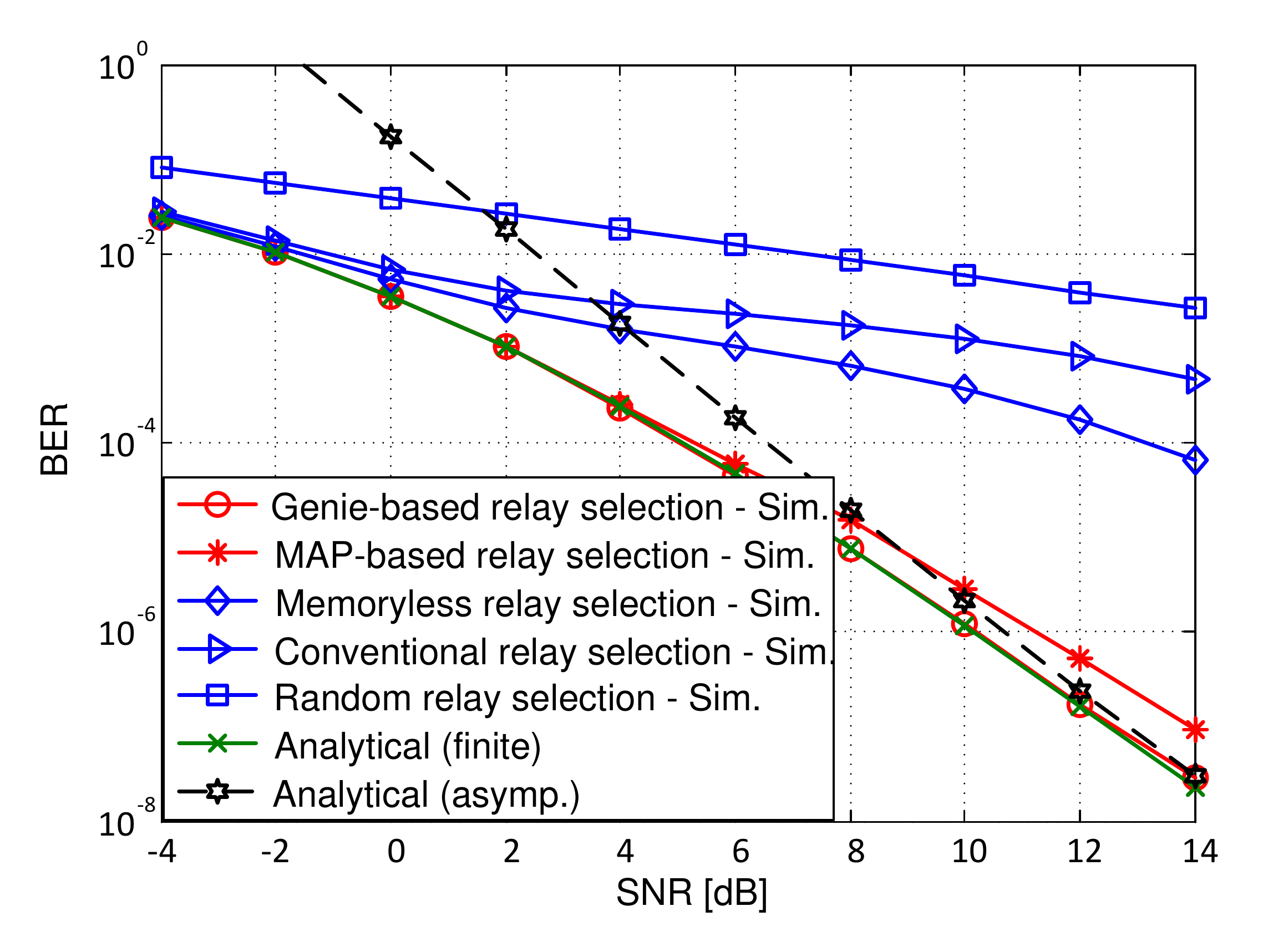}\\
  \caption{BER performances at the $N$'th best relay for various relay selection schemes with $M=5$ relays over Rayleigh faded TSMG channels. A system involving an uncoded transmission and a BPSK modulation is considered.}\label{ber_bestrelay}
\end{figure}

Fig.~\ref{ber_bestrelay} depicts both the analytical and simulated BER performance at the selected relay, assuming different relay selection protocols. The derived BER expression in (\ref{ber_sr_finite}) is used to evaluate the exact analytical result and its asymptotic performance is evaluated using (\ref{ber_sr_final}). The simulated BER performances are obtained by averaging the error rate over $10^5$ frames with $10^4$ samples for every frame. Fig.~\ref{ber_bestrelay} shows that the simulation results for the genie-aided selection perfectly match the analytical results. Also, the derived asymptotic error rate expression accurately predicts the performance for sufficiently high SNR. However, the genie detection is practically infeasible. Interestingly, we also remark that the performance of the proposed $N$'th BRS scheme, employing MAP-detection, almost approaches the genie-aided case and provides a significant performance gain over the other schemes. Obviously, this comes at the cost of higher complexity required for the implementation of the forward-backward algorithm. Hence, when the noise memory is exploited in the relay selection process through MAP detection, we achieve a significant performance gain. Finally, the proposed simpler memoryless algorithm still exhibits a better performance than conventional BRS schemes, by taking into account the partial IN statistics in the selection process.
\begin{figure}[!t]
  \centering
  \includegraphics[width=\columnwidth]{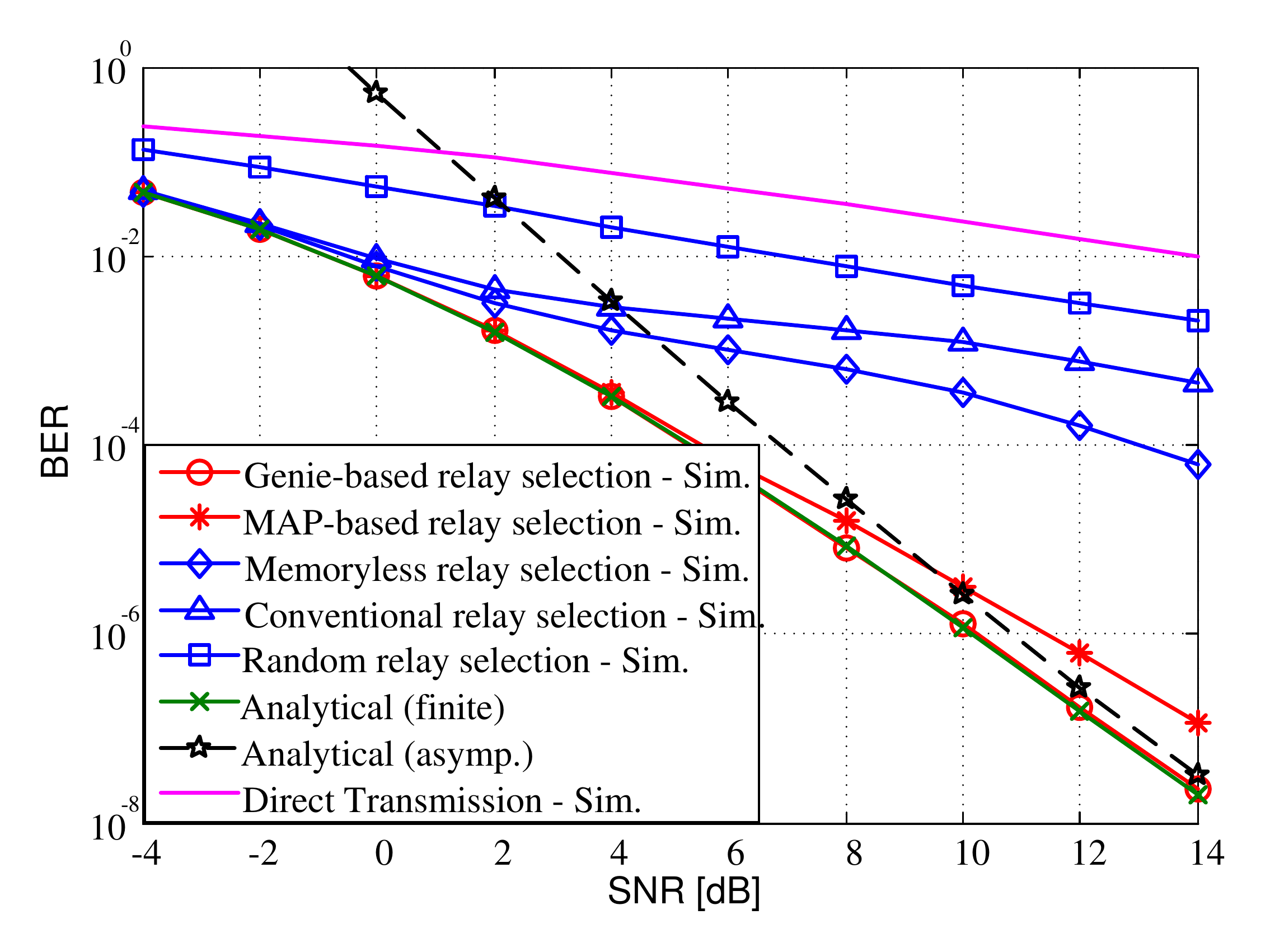}\\
  \caption{End-to-end BER performances of various $N$'th BRS schemes with $M=5$ relays over Rayleigh faded TSMG channels. A system involving an uncoded transmission and a BPSK modulation is considered.}\label{ber_resec_ete}
\end{figure}

Fig.~\ref{ber_resec_ete} shows the end-to-end analytical and simulated BER performances for the proposed scenario. The analytical BERs are evaluated using (\ref{ber_atdestination}) and (\ref{ber_atdestination_imp}), respectively for both cases of when the selected relay is in good state or in bad state. As a benchmark, we also include the performance of direct transmission (DT) over Rayleigh faded AWGN channel. From Fig.~\ref{ber_resec_ete}, we observe that, the end-to-end analytical BER corresponds to the simulation results for genie selection and its asymptotic performance reflects the exact performance for sufficiently high SNR. This further confirms that the proposed MAP-based $N$'th BRS scheme efficiently decreases the effects of IN which significantly improves the system performance compared to conventional schemes. Moreover, even when subjected to IN, CR outperforms DT irrespective of the relay selection process, however, the amount of improvement depends on the process.
\begin{figure}[!t]
  \centering
  \includegraphics[width=\columnwidth]{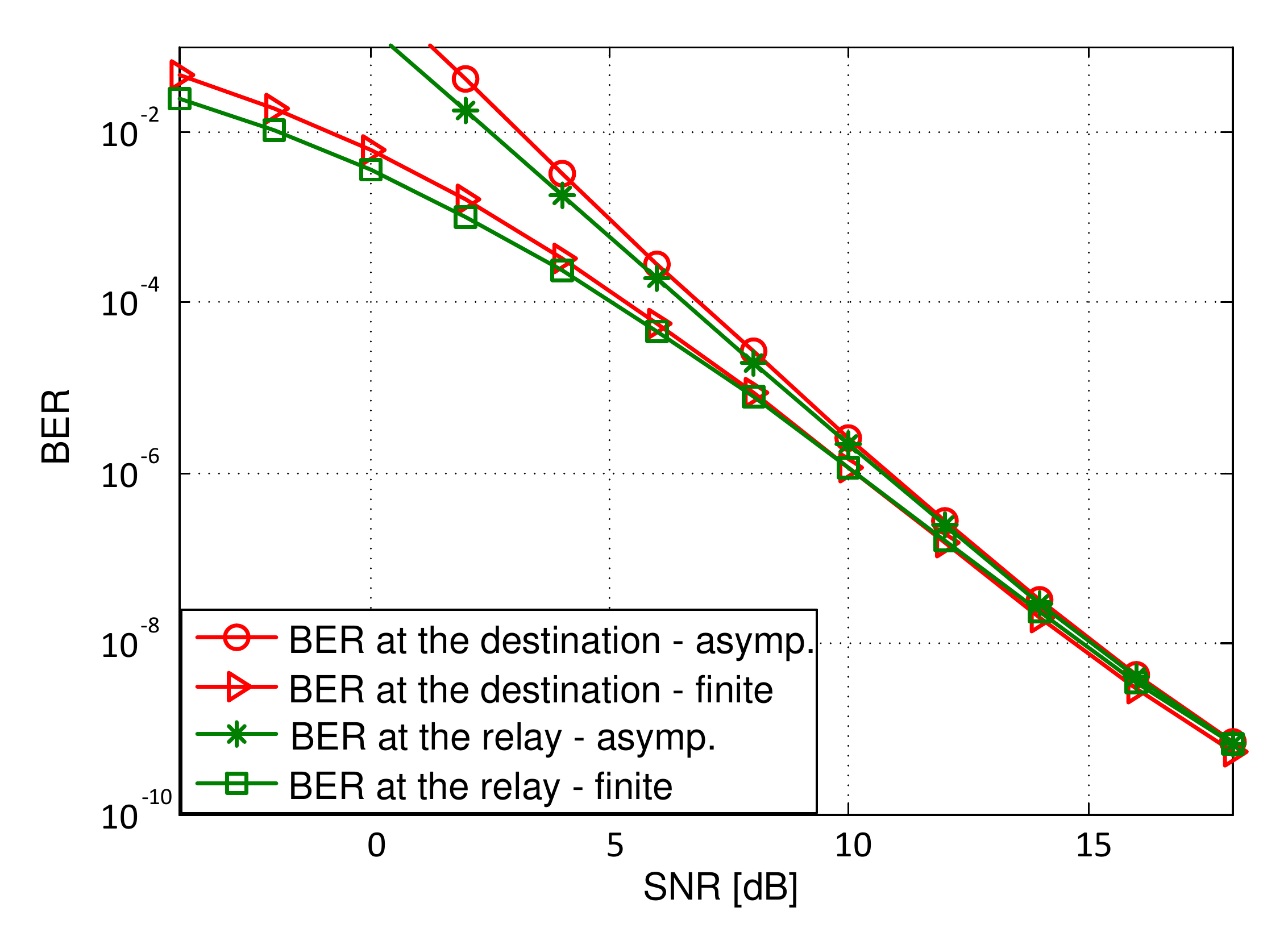}\\
  \caption{Analytical asymptotic and finite BER performances at the $N$'th best relay and at the destination with $M=5$ relays over Rayleigh faded TSMG channels.}\label{ber_finite_vs_asymp}
\end{figure}

To circumvent the burden of obtaining the time-consuming simulation results in the BER range of $10^{-8}-10^{-10}$, Fig.~\ref{ber_finite_vs_asymp} illustrates the analytical BER performances only. From Fig.~\ref{ber_finite_vs_asymp}, it is obvious that the asymptotic performance truly reflects the finite SNR BER performance for sufficiently high SNR. Therefore, we can check the diversity order of each relay selection scheme by taking the slope of BER performances shown in Fig.~\ref{ber_resec_ete} \cite{al2009cooperative}. It is verified that the obtained diversity orders of MAP-based, memoryless, conventional, and random relay selection schemes are respectively, $5.9$, $3.85$, $3.3$, and $2.9$. Hence, the proposed MAP-based $N$'th BRS scheme achieves the full diversity order.
\begin{figure}[!t]
  \centering
  \includegraphics[width=\columnwidth]{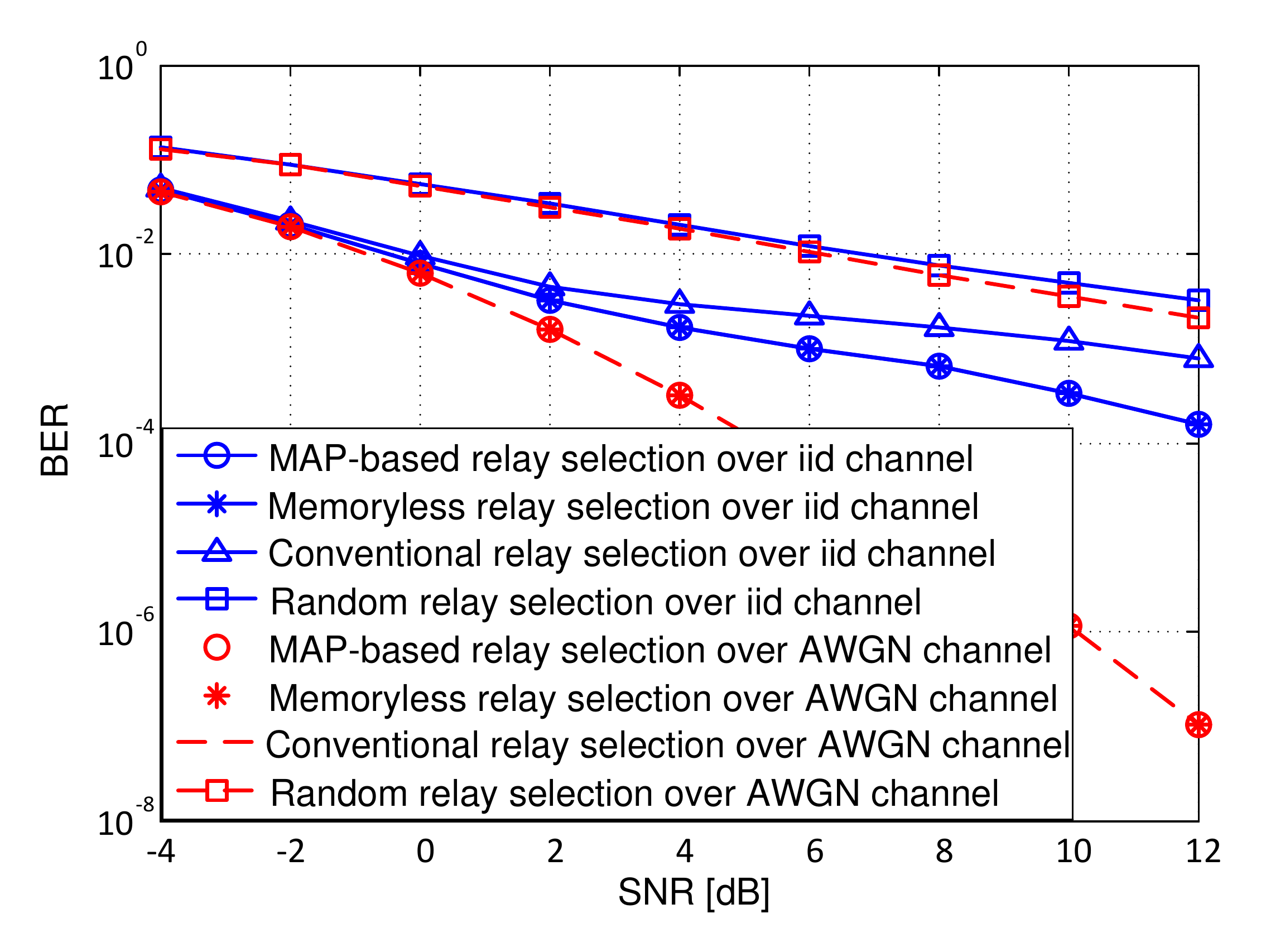}\\
  \caption{End-to-end BER performances of various $N$'th BRS schemes with $M=5$ relays. A system involving an uncoded transmission and BPSK modulation is considered. It is assumed that $p_B=0.01$ with $\mu=1$, $\rho=100$ for the i.i.d. channel, and $\mu=1$, $\rho=1$ for the AWGN channel.}\label{ber_resec_ete_IID_AWGN}
\end{figure}

Fig.~\ref{ber_resec_ete_IID_AWGN} presents the simulated end-to-end BER performances of the proposed relay selection protocols in case of memoryless impulsive and AWGN noise scenario. From Fig.~\ref{ber_resec_ete_IID_AWGN}, it is observed that the MAP-based and memoryless relay selection schemes show the same performance, when the noise memory is reduced from $\mu=100$ in Fig.~\ref{ber_resec_ete} to $\mu=1$, i.e., in case of memoryless impulsive noise.
This confirms that, for memoryless IN, the optimal MAP detector simplifies to the sample-by-sample detector. Again, the conventional relay selection achieves the worst performance in these impulsive scenarios. We further show the corresponding results in case of Gaussian channel. Obviously, the three relay selection schemes provide the same performance in this case.
\begin{figure}[!t]
  \centering
  \includegraphics[width=\columnwidth]{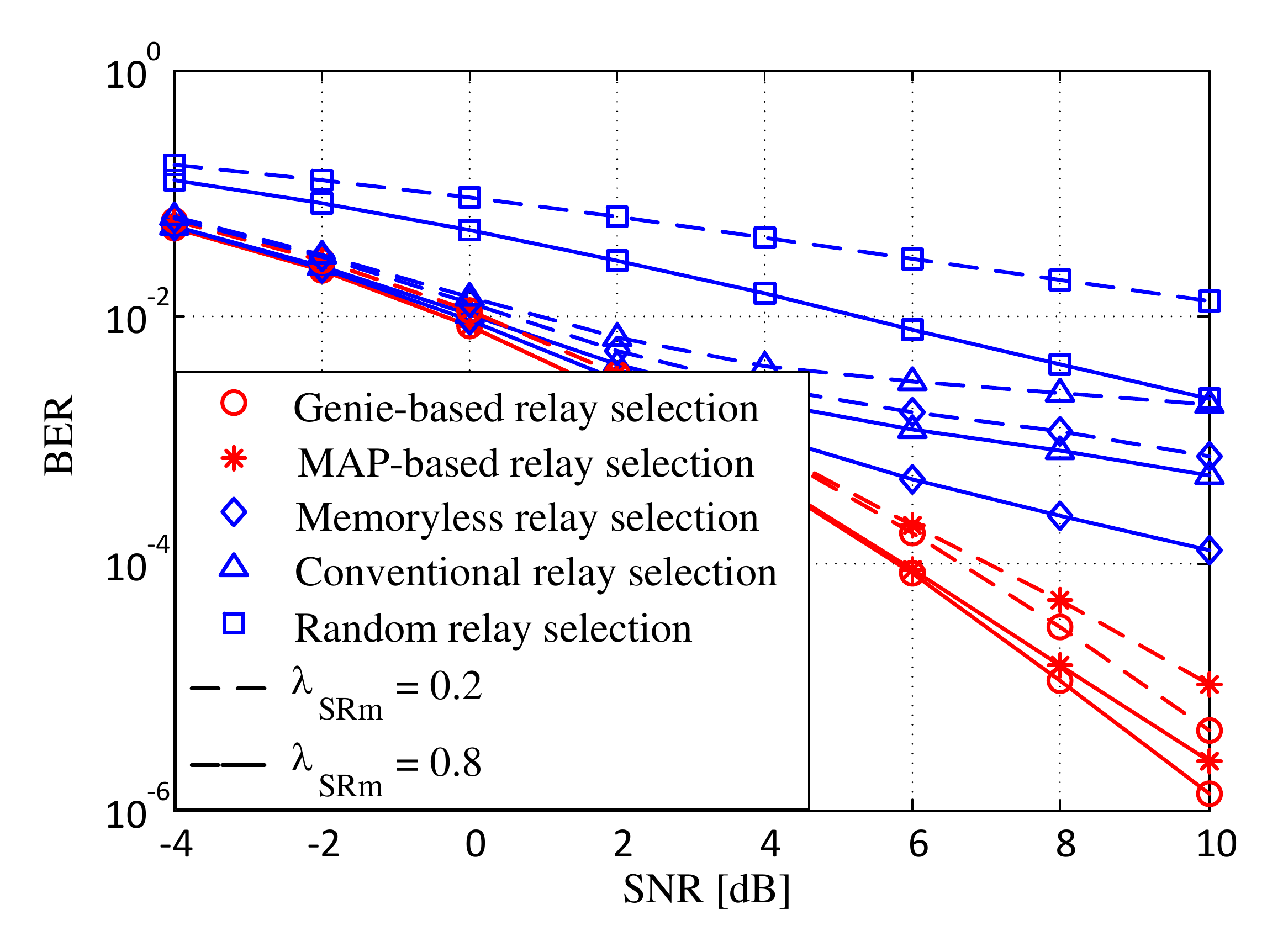}\\
  \caption{End-to-end BER performances of various $N$'th BRS schemes for various best relay positions. A system involving an uncoded transmission with $M=5$ relays over Rayleigh faded TSMG channels and a BPSK modulation is considered.}\label{ber_ete_relay_position}
\end{figure}

In order to illustrate the effect of best relay location, we demonstrate in Fig.~\ref{ber_ete_relay_position} the performance of the considered relay selection schemes for asymmetric network scenarios with $\lambda_{SR_m}=0.2$ and $\lambda_{SR_m}=0.8$. We observe from Fig.~\ref{ber_ete_relay_position} that the performance of opportunistic relaying degrades if the best relay is moved from the source to the destination irrespective of the relay selection process. It turns out that the best relay being closer to the source is more rewarding than closer to the destination.
\begin{figure}[!t]
  \centering
  \includegraphics[width=\columnwidth]{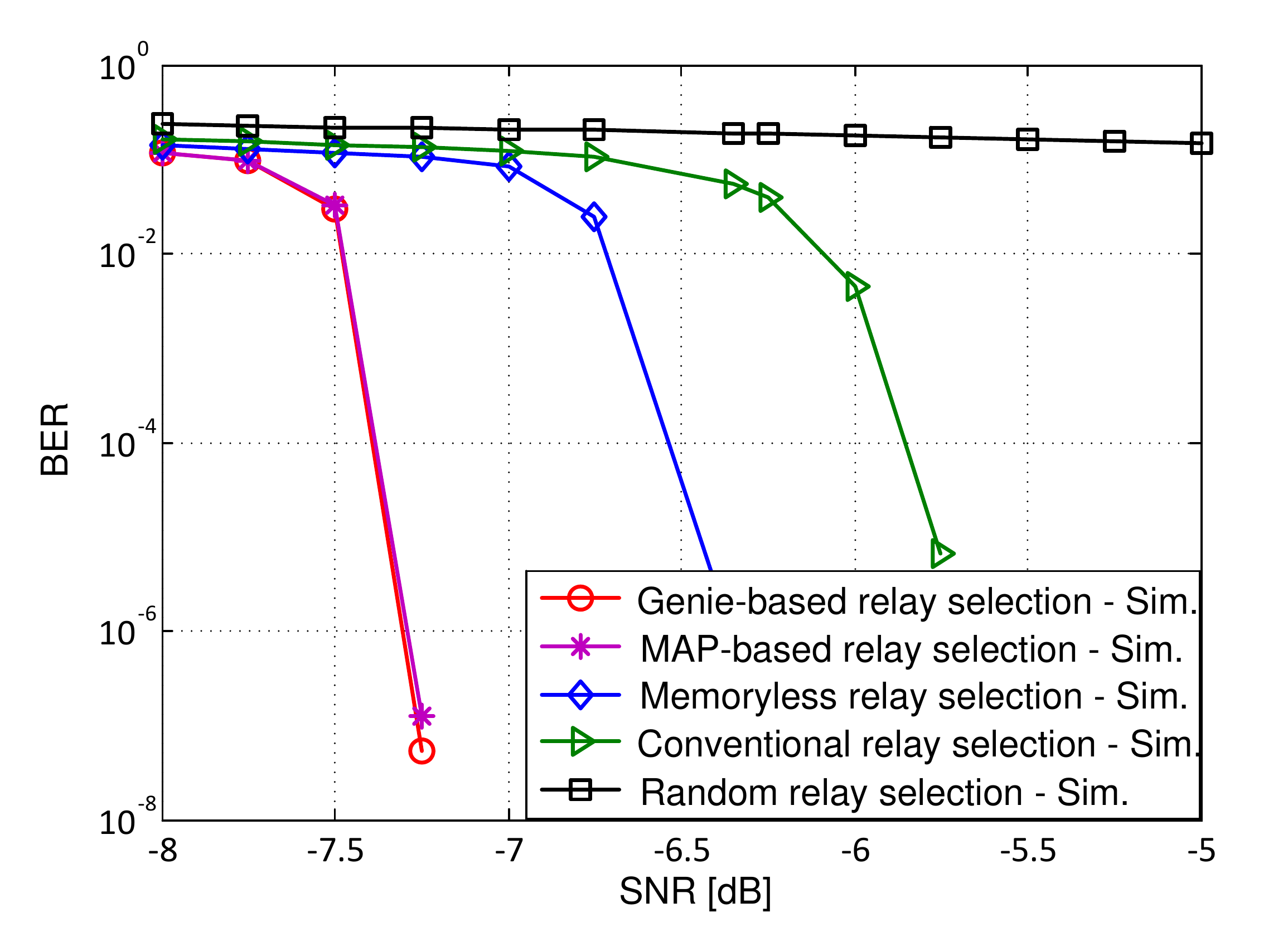}\\
  \caption{BER performances at the $N$'th best relay of various BRS schemes with $M=5$ relays over Rayleigh faded TSMG channels. A system involving an  LDPC coded transmission and BPSK modulation is considered.}\label{ber_ete_coded}
\end{figure}

Furthermore, we also investigate the performance of the proposed relay selection schemes under coded transmission. It is interesting to evaluate how much gain does the proposed MAP-based relay selection scheme provide over the other schemes for systems employing powerful channel codes such as low-density parity check (LDPC) codes. In Fig.~\ref{ber_ete_coded}, we show the simulated BERs at the selected relay for various relay selection schemes under LDPC coded transmission. At the transmitter, a frame of equally likely $32,400$ information bits is first encoded with the code rate of $1/2$ and then mapped to a BPSK modulation sequence. For LDPC decoding, we set the number of iterations to $50$. As expected, from Fig.~\ref{ber_ete_coded}, we remark that similar to uncoded transmission, the proposed MAP-based $N$'th BRS scheme provides a significant performance gain over the other schemes.
\begin{figure}[!t]
  \centering
  \includegraphics[width=\columnwidth]{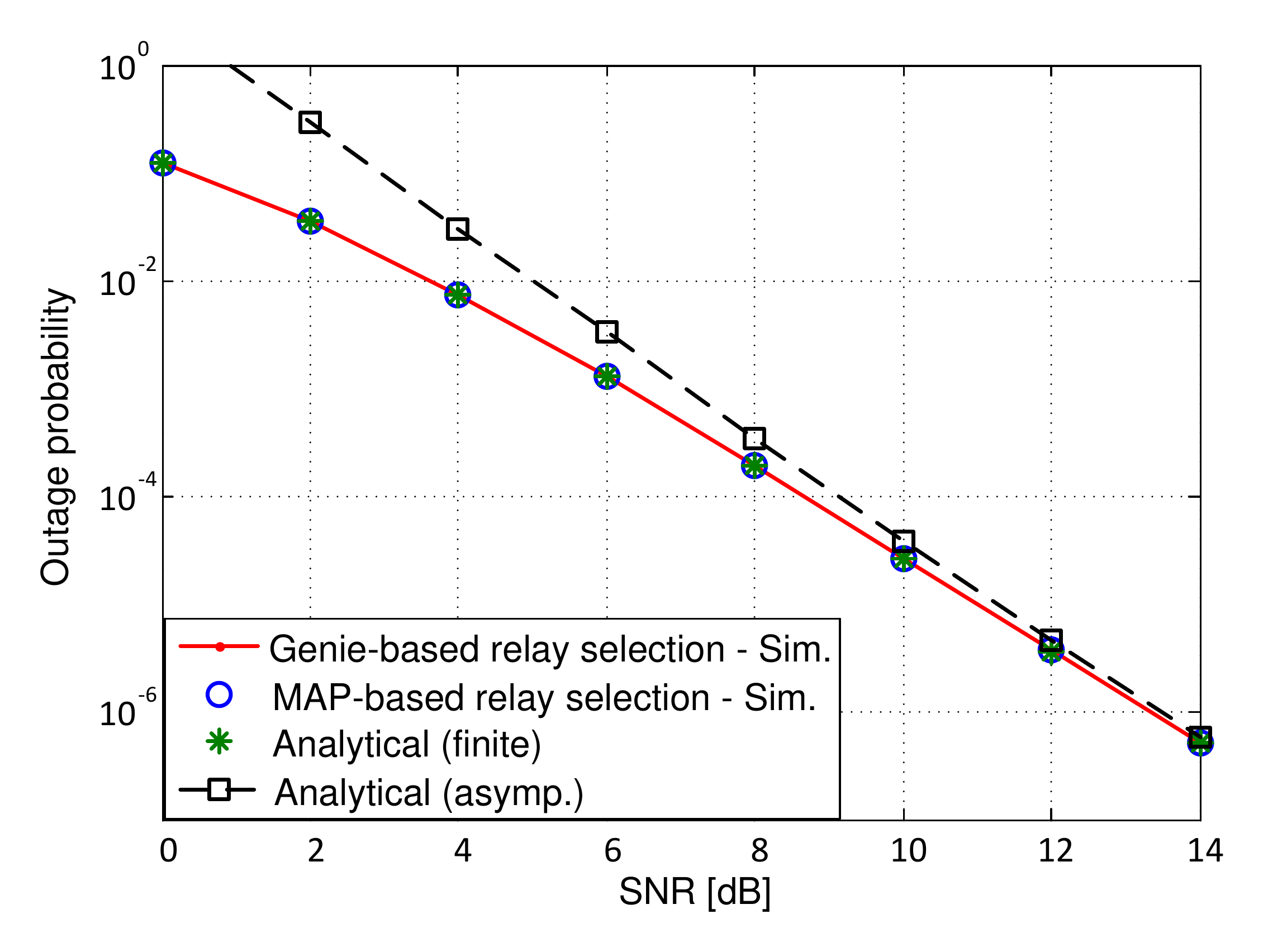}\\
  \caption{Outage performances at the $N$'th best relay of various relay selection schemes with $M=5$ relays over Rayleigh faded TSMG channels. A system involving an uncoded transmission and a BPSK modulation is considered.}\label{outage_bestrelay}
\end{figure}
\begin{figure}[!t]
  \centering
  \includegraphics[width=\columnwidth]{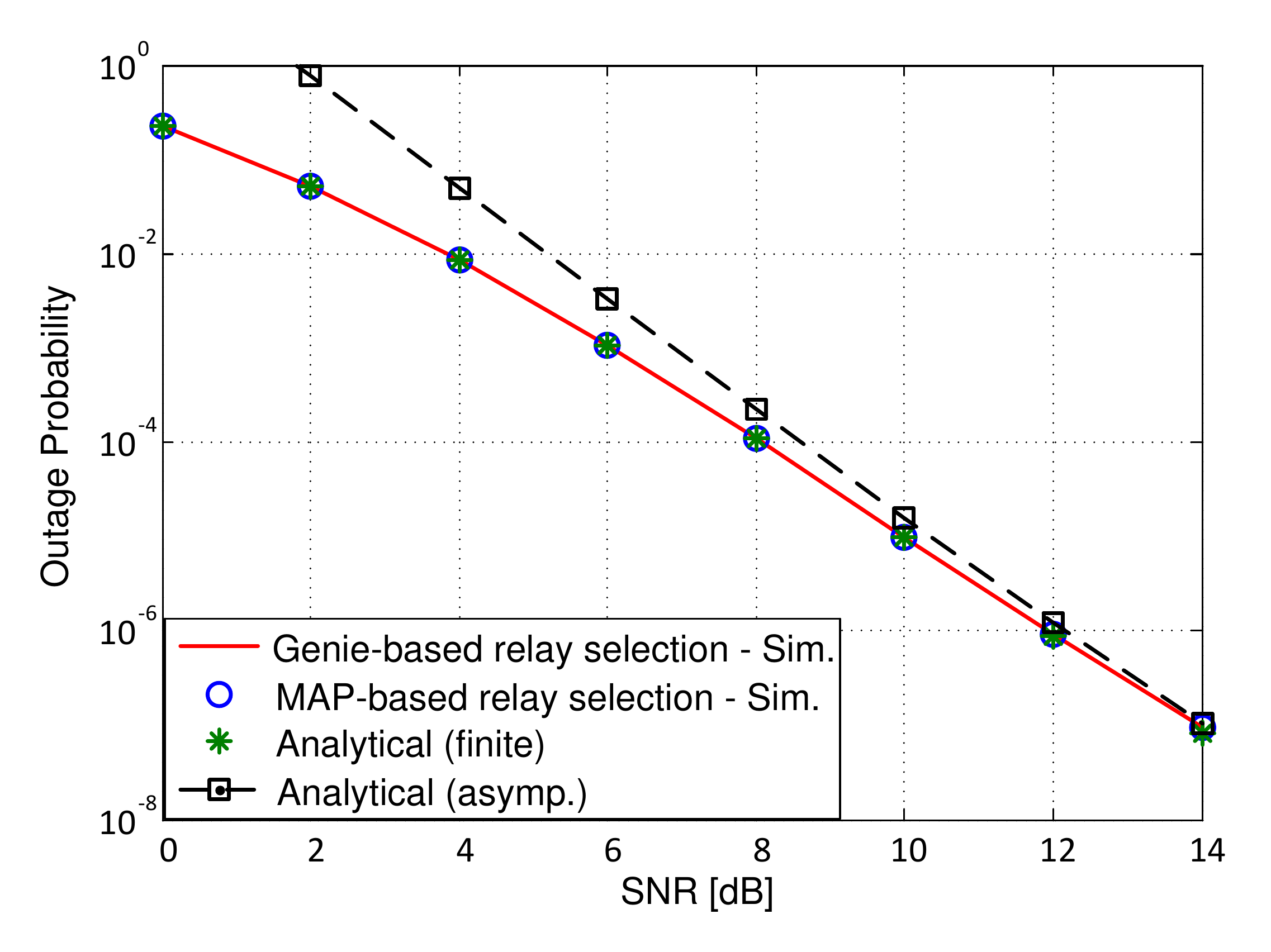}\\
  \caption{End-to-end outage performances of various $N$'th BRS schemes with $M=5$ relays over Rayleigh faded TSMG channels. A system involving an uncoded transmission and a BPSK modulation is considered.}\label{outage_resec_ete}
\end{figure}

Figures \ref{outage_bestrelay} and \ref{outage_resec_ete} depict the outage probability and the corresponding asymptotic curves at the selected relay as well as the destination for a targeted data rate $R=1$ bits/s/Hz. It is observed from figures \ref{outage_bestrelay} and \ref{outage_resec_ete} that the derived analytical outage performances provide an exact match to the simulation results for the genie-aided scheme. It also observed that the MAP-based BRS scheme performs exactly as the genie-aided scheme. Therefore, the MAP-based relay selection criterion is the most suitable one for bursty IN environments as it has been designed according to the statistical behavior of the noise. In addition, it achieves the full diversity order of $M+1$ as shown in Fig.~\ref{outage_resec_ete}.
\section{Conclusion}\label{conclusion}
In this article, we have investigated the performance of some conventional relay selection protocols for DF CR over Rayleigh faded bursty IN channels and have proposed an improved approach for relay selection. The proposed method avoids the use of error detection methods at the relay nodes and is based on both the channel state information of source-relay and relay-destination links, and the state of the IN that affect those links. We provided closed-form expressions for the PDF of the received SNR at the $N$'th best relay as well as at the destination under both cases of finite SNR and asymptotic analysis. As a consequence, these PDFs are used to derive closed-form expressions for the end-to-end BER, as well as the outage probability, facilitating the achievement of the diversity order of the scheme. Simulation results confirmed the accuracy of the proposed asymptotic and finite SNR analysis. From the obtained results, it is verified that our proposed MAP-based $N$'th BRS scheme outperforms the conventional schemes optimized for the Gaussian case, and which cannot take into account the IN memory.
\appendices
\section{Derivation of the Marginal PDF: Asymptotic case}
In the high SNR regime, it is assumed that $1-e^{-x}\doteq x$. Then, from (\ref{exact_value}), we have
\begin{eqnarray}
I_1&\doteq&M{{M-1} \choose {N-1}}\frac{1}{\overline{\gamma}_{SR_m}\overline{\gamma}_{R_mD}}e^{-\frac{x}{\overline{\gamma}_{SR_m}}}\left(\frac{1}{\overline{\gamma}_{a}}\right)^{M-N}\nonumber\\
&&\times \int_{z=0}^x z^{M-N}e^{-\frac{z}{\overline{\gamma}_{R_mD}}}dz,\label{marginal_high_sr}
\end{eqnarray}
Assuming $\frac{z}{\overline{\gamma}_{R_mD}}=q$, (\ref{marginal_high_sr}) can be rearranged as
\begin{eqnarray}
I_1&\!\!\!=\!\!\!&\frac{C_M}{\overline{\gamma}_{SR_m}\overline{\gamma}_{R_mD}}e^{-\frac{x}{\overline{\gamma}_{SR_m}}}\!\left(\frac{1}{\overline{\gamma}_{a}}\!\right)^{M-N}\!\!\!\!\int_{z=0}^{\frac{x}{\overline{\gamma}_{R_mD}}}\!\!q^{M-N}e^{-q}dq,\nonumber\\
&\!\!\!=\!\!\!&\frac{C_M}{\overline{\gamma}_{SR_m}}e^{-\frac{x}{\overline{\gamma}_{SR_m}}}\left(\frac{1}{\overline{\gamma}_{a}}\right)^{M-N}\!\!\!\!\gamma(M-N+1,\frac{x}{\overline{\gamma}_{R_mD}}),
\end{eqnarray}
where $\gamma(a,b)$ is the lower incomplete gamma function. At high SNR scenario, $x\approx 0$ and $I_1$ can be neglected. Again,
\begin{eqnarray}
I_2&=&\frac{C_M}{\overline{\gamma}_{SR_m}\overline{\gamma}_{R_mD}}\left(\frac{x}{\overline{\gamma}_{a}}\right)^{M-N}e^{-\frac{x}{\overline{\gamma}_{SR_m}}}\int_{z=x}^\infty e^{-\frac{z}{\overline{\gamma}_{R_mD}}}dz, \nonumber\\
&=& M{{M-1} \choose {N-1}}\frac{1}{\overline{\gamma}_{SR_m}}\left(\frac{x}{\overline{\gamma}_{a}}\right)^{M-N}e^{-\frac{x}{\overline{\gamma}_{a}}},\label{marginal_pdf_exact}\\
&\approx& M{{M-1} \choose {N-1}}\frac{1}{\overline{\gamma}_{SR_m}}\left(\frac{1}{\overline{\gamma}_{a}}\right)^{M-N}x^{M-N}.
\end{eqnarray}
Substituting $I_1$ and $I_2$ in (\ref{exact_value}), yields (\ref{marginal_PDF_sr}).
\ifCLASSOPTIONcaptionsoff
  \newpage
\fi



%
%
\bibliographystyle{ieeetr}
\bibliography{book_ieee_TVT}

\begin{thebibliography}{10}

\bibitem{nosratinia2004cooperative}
A.~Nosratinia, T.~E. Hunter, and A.~Hedayat, ``Cooperative communication in
  wireless networks,'' {\em IEEE Commun. Mag.}, vol.~42, no.~10, pp.~74--80,
  2004.

\bibitem{laneman2004cooperative}
J.~N. Laneman, D.~N. Tse, and G.~W. Wornell, ``Cooperative diversity in
  wireless networks: efficient protocols and outage behavior,'' {\em IEEE
  Trans. Inf. Theory}, vol.~50, no.~12, pp.~3062--3080, 2004.

\bibitem{laneman2003distributed}
J.~N. Laneman and G.~W. Wornell, ``Distributed space-time-coded protocols for
  exploiting cooperative diversity in wireless networks,'' {\em IEEE Trans.
  Inf. Theory}, vol.~49, no.~10, pp.~2415--2425, 2003.

\bibitem{bletsas2006simple}
A.~Bletsas, A.~Khisti, D.~P. Reed, and A.~Lippman, ``A simple cooperative
  diversity method based on network path selection,'' {\em IEEE J. Sel. Areas
  Commun.}, vol.~24, no.~3, pp.~659--672, 2006.

\bibitem{ibrahim2008cooperative}
A.~S. Ibrahim, A.~K. Sadek, W.~Su, and K.~R. Liu, ``Cooperative communications
  with relay-selection: when to cooperate and whom to cooperate with?,'' {\em
  IEEE Trans. Wireless Commun.}, vol.~7, no.~7, pp.~2814--2827, 2008.

\bibitem{fareed2009relay}
M.~M. Fareed and M.~Uysal, ``On relay selection for decode-and-forward
  relaying,'' {\em IEEE Trans. Wireless Commun.}, vol.~8, no.~7,
  pp.~3341--3346, 2009.

\bibitem{tourki2013new}
K.~Tourki, H.-C. Yang, M.-S. Alouini, and K.~A. Qaraqe, ``New results on
  performance analysis of opportunistic regenerative relaying,'' {\em Physical
  Communication}, vol.~9, pp.~97--111, 2013.

\bibitem{ikki2010performance}
S.~S. Ikki and M.~H. Ahmed, ``On the performance of cooperative-diversity
  networks with the ${N^{th}}$ best-relay selection scheme,'' {\em IEEE Trans.
  Commun.}, vol.~58, no.~11, pp.~3062--3069, 2010.

\bibitem{al2018asymptotic}
Y.~H. Al-Badarneh, C.~N. Georghiades, and M.-S. Alouini, ``Asymptotic
  performance analysis of the $k$-th best link selection over wireless fading
  channels: An extreme value theory approach,'' {\em IEEE Trans. Veh.
  Technol.}, vol.~67, no.~7, pp.~6652--6657, 2018.

\bibitem{zhang2015performance}
X.~Zhang, Y.~Zhang, Z.~Yan, J.~Xing, and W.~Wang, ``Performance analysis of
  cognitive relay networks over {Nakagami}-$m$ fading channels,'' {\em IEEE J.
  Sel. Areas Commun.}, vol.~33, no.~5, pp.~865--877, 2015.

\bibitem{al2019asymptotic}
Y.~H. Al-Badarneh, C.~N. Georghiades, and M.-S. Alouini, ``On the asymptotic
  throughput of the $ k $-th best secondary user selection in cognitive radio
  systems,'' in {\em Proc. IEEE Veh. Technol. Conf.}, pp.~1--5, IEEE, 2019.

\bibitem{zhang2018secrecy}
J.~Zhang, G.~Pan, and Y.~Xie, ``Secrecy analysis of wireless-powered
  multi-antenna relaying system with nonlinear energy harvesters and imperfect
  {CSI},'' {\em IEEE Trans. Green Commun. Net.}, vol.~2, no.~2, pp.~460--470,
  2018.

\bibitem{davidorder}
H.~A. David and H.~N. Nagaraja, {\em Order Statistics}.
\newblock Hoboken, NJ, USA: Wiley, 2003.

\bibitem{middleton1977statistical}
D.~Middleton, ``Statistical-physical models of electromagnetic interference,''
  {\em IEEE Trans. Electromagn. Compat.}, vol.~EMC-19, no.~3, pp.~106--127,
  1977.

\bibitem{sacuto2014wide}
F.~Sacuto, F.~Labeau, and B.~L. Agba, ``Wide band time-correlated model for
  wireless communications under impulsive noise within power substation,'' {\em
  IEEE Trans. Wireless Commun.}, vol.~13, no.~3, pp.~1449--1461, 2014.

\bibitem{zimmermann2002analysis}
M.~Zimmermann and K.~Dostert, ``Analysis and modeling of impulsive noise in
  broad-band powerline communications,'' {\em IEEE Trans. Electromagn.
  Compat.}, vol.~44, no.~1, pp.~249--258, 2002.

\bibitem{ndomarkov}
G.~Ndo, F.~Labeau, and M.~Kassouf, ``A {Markov-Middleton} model for bursty
  impulsive noise: modeling and receiver design,'' {\em IEEE Trans. Power
  Del.}, vol.~28, no.~4, pp.~2317--2325, 2013.

\bibitem{shongwe2015study}
T.~Shongwe, A.~Han~Vinck, and H.~C. Ferreira, ``A study on impulse noise and
  its models,'' {\em SAIEE Africa Res. J.}, vol.~106, no.~3, pp.~119--131,
  2015.

\bibitem{agba2019impulsive}
B.~L. Agba, F.~Sacuto, M.~Au, F.~Labeau, and F.~Gagnon, ``Impulsive noise
  measurements,'' in {\em Wireless Communications for Power Substations: RF
  Characterization and Modeling}, pp.~35--68, Springer, 2019.

\bibitem{blackard1993measurements}
K.~L. Blackard, T.~S. Rappaport, and C.~W. Bostian, ``Measurements and models
  of radio frequency impulsive noise for indoor wireless communications,'' {\em
  IEEE J. Sel. Areas Commun.}, vol.~11, no.~7, pp.~991--1001, 1993.

\bibitem{cheffena2012industrial}
M.~Cheffena, ``Industrial wireless sensor networks: channel modeling and
  performance evaluation,'' {\em EURASIP J. Wireless Commun. Net.}, vol.~297,
  no.~1, pp.~1--8, 2012.

\bibitem{asiyo2017analysis}
M.~O. Asiyo and T.~J. Afullo, ``Analysis of bursty impulsive noise in
  low-voltage indoor power line communication channels: local scaling
  behaviour,'' {\em SAIEE Africa Res. J.}, vol.~108, no.~3, pp.~98--107, 2017.

\bibitem{bai2017discrete}
T.~Bai, H.~Zhang, R.~Zhang, L.-L. Yang, A.~F. Al~Rawi, J.~Zhang, and L.~Hanzo,
  ``Discrete multi-tone digital subscriber loop performance in the face of
  impulsive noise,'' {\em IEEE Access}, vol.~5, pp.~10478--10495, 2017.

\bibitem{qian2018performance}
Y.~Qian, J.~Li, Y.~Zhang, and D.~N.~K. Jayakody, ``Performance analysis of an
  opportunistic relaying power line communication systems,'' {\em IEEE Systems
  J.}, vol.~12, no.~4, pp.~3865--3868, 2018.

\bibitem{krikidis2009max}
I.~Krikidis, J.~S. Thompson, S.~McLaughlin, and N.~Goertz, ``Max-min relay
  selection for legacy amplify-and-forward systems with interference,'' {\em
  IEEE Trans. Wireless Commun.}, vol.~8, no.~6, pp.~3016--3027, 2009.

\bibitem{ahmed2012relay}
I.~Ahmed, A.~Nasri, D.~S. Michalopoulos, R.~Schober, and R.~K. Mallik, ``Relay
  subset selection and fair power allocation for best and partial relay
  selection in generic noise and interference,'' {\em IEEE Trans. Wireless
  Commun.}, vol.~11, no.~5, pp.~1828--1839, 2012.

\bibitem{alam2016performance}
M.~S. Alam, F.~Labeau, and G.~Kaddoum, ``Performance analysis of {DF}
  cooperative relaying over bursty impulsive noise channel,'' {\em IEEE Trans.
  Commun.}, vol.~64, no.~7, pp.~2848--2859, 2016.

\bibitem{alam2016effect}
M.~S. Alam and F.~Labeau, ``Effect of bursty impulsive noise on the performance
  of multi-relay {DF} cooperative relaying scheme,'' in {\em Proc. IEEE
  Vehicular Technology Conf.}, pp.~1--5, IEEE, 2016.

\bibitem{alam2016relay}
M.~S. Alam and F.~Labeau, ``On relay selection in bursty impulsive noise
  channel,'' in {\em Proc. IEEE Wireless Commun. Networking Conf.}, pp.~1--6,
  IEEE, 2016.

\bibitem{fertonani2009reliable}
D.~Fertonani and G.~Colavolpe, ``On reliable communications over channels
  impaired by bursty impulse noise,'' {\em IEEE Trans. Commun.}, vol.~57,
  no.~7, pp.~2024--2030, 2009.

\bibitem{mitra2010convolutionally}
J.~Mitra and L.~Lampe, ``Convolutionally coded transmission over
  {Markov-Gaussian} channels: analysis and decoding metrics,'' {\em IEEE Trans.
  Commun.}, vol.~58, no.~7, pp.~1939--1949, 2010.

\bibitem{bahl1974optimal}
L.~Bahl, J.~Cocke, F.~Jelinek, and J.~Raviv, ``Optimal decoding of linear codes
  for minimizing symbol error rate,'' {\em IEEE Trans. Inf. Theory}, vol.~20,
  no.~2, pp.~284--287, 1974.

\bibitem{wang2007high}
T.~Wang, A.~Cano, G.~B. Giannakis, and J.~N. Laneman, ``High-performance
  cooperative demodulation with decode-and-forward relays,'' {\em IEEE Trans.
  Commun.}, vol.~55, no.~7, pp.~1427--1438, 2007.

\bibitem{ghosh1996analysis}
M.~Ghosh, ``Analysis of the effect of impulse noise on multicarrier and single
  carrier {QAM} systems,'' {\em IEEE Trans. Commun.}, vol.~44, no.~2,
  pp.~145--147, 1996.

\bibitem{fertonani2007reduced}
D.~Fertonani, A.~Barbieri, and G.~Colavolpe, ``Reduced-complexity {BCJR}
  algorithm for turbo equalization,'' {\em IEEE Trans. Commun.}, vol.~55,
  no.~12, pp.~2279--2287, 2007.

\bibitem{papoulis2002probability}
A.~Papoulis and S.~U. Pillai, ``Probability, random variables and stochastic
  processes,'' {\em New York, NY, McGraw-Hill Education}, 2002.

\bibitem{proakis2001digital}
J.~G. Proakis, {\em Digital Communications, 4th edition}.
\newblock McGraw-Hill, 2001.

\bibitem{goldsmith2005wireless}
A.~Goldsmith, {\em Wireless communications}.
\newblock Cambridge university press, 2005.

\bibitem{al2009cooperative}
S.~Al-Dharrab and M.~Uysal, ``Cooperative diversity in the presence of
  impulsive noise,'' {\em IEEE Trans. Wireless Commun.}, vol.~8, no.~9,
  pp.~4730--4739, 2009.

\end{thebibliography}

\end{document}